\documentclass[epsfig,12pt]{article}
\usepackage{makeidx}
\usepackage{amsmath}
\usepackage{amsfonts}
\usepackage{amssymb}
\usepackage{graphicx}

\input epsf.sty
\makeatletter \@addtoreset{equation}{section}
\renewcommand{\theequation}{\thesection.\arabic{equation}}
\input{tcilatex}
\begin{document}

\title{\vspace{-2cm}\rightline{\mbox{\small {\bf LPHE-MS-11-01  /
CPM-11-01}}} \bigskip \bigskip \textbf{Four Dimensional Graphene}}
\author{L.B Drissi $^{1,2}$, E.H Saidi $^{2,3,4}$, M. Bousmina $^{2}$%
\bigskip  \\
$^{1}${\small International Centre for Theoretical Physics, ICTP, Trieste,
Italy,}\\
$^{{\small 2}}${\small MAScIR, Institute of Nanomaterials and
Nanotechnology, ENSET, Rabat, Morocco}\\
$^{{\small 3}}${\small LPHE-Modeling and Simulation, Facult\'{e} des
Sciences, Rabat, Morocco}\\
$^{{\small 4}}${\small Centre of Physics and Mathematics, CPM- CNESTEN,
Morocco}}
\maketitle

\begin{abstract}
Mimicking pristine \emph{2D} graphene, we revisit the BBTW model \ for \emph{%
4D }lattice QCD given in [\emph{P. F. Bedaque et al. Phys. Rev. D78 (2008)
017502}] by using the hidden $SU\left( 5\right) $ symmetry of the \emph{4D}
hyperdiamond lattice $\mathcal{H}_{4}$. We first study the link between the $%
\mathcal{H}_{4}$ and $SU\left( 5\right) $; then we refine the BBTW \emph{4D}
lattice action by using the weight vectors $\mathbf{\lambda }_{1},$ $\mathbf{%
\lambda }_{2},$ $\mathbf{\lambda }_{3},$ $\mathbf{\lambda }_{4},$ $\mathbf{%
\lambda }_{5}$ of the \emph{5}-dimensional representation of $SU\left(
5\right) $ satisfying $\sum_{i}\mathbf{\lambda }_{i}=0$. After that we study
explicitly the solutions of the zeros of the Dirac operator $\mathcal{D}$ in
terms of the SU$\left( 5\right) $ simple roots $\mathbf{\alpha }_{1},$ $%
\mathbf{\alpha }_{2},$ $\mathbf{\alpha }_{3},$ $\mathbf{\alpha }_{4}$
generating $\mathcal{H}_{4}$; and its fundamental weights $\omega _{1},$ $%
\omega _{2},$ $\omega _{3},$ $\omega _{4}$ which generate the reciprocal
lattice $\mathcal{H}_{4}^{\ast }$. It is shown, amongst others, that these
zeros live at the sites of $\mathcal{H}_{4}^{\ast }$; and the continuous
limit $\mathcal{D}$ is given by $\frac{id\sqrt{5}}{2}$ $\gamma ^{\mu }%
\mathbf{k}_{\mu }$ with $d,$ $\gamma ^{\mu }$ and $\mathbf{k}_{\mu }$
standing respectively for the lattice parameter of $\mathcal{H}_{4}$, the
usual \emph{4} Dirac matrices and the \emph{4D} wave vector. Other features
such as differences with BBTW model as well as the link between the Dirac
operator following from our construction and the one suggested by Creutz
using quaternions, are also given{.}

\end{abstract}


\section{Introduction}

In the last few years, there have been attempts to extend results on the
relativistic electron system on a \emph{2D} honeycomb (graphene) \textrm{%
\cite{A1,A2,A3}} to a \emph{4D} honeycomb lattice (called \emph{4D }%
hyperdiamond, denoted below as $\mathcal{H}_{4}$) and apply it to the
lattice QCD simulations \textrm{\cite{B1}-\cite{B6}}. These attempts try to
construct Dirac fermion on $\mathcal{H}_{4}$ by keeping all desirable
properties; in particular locality, chiral symmetry and the minimal number
of fermion doublings \textrm{\cite{B1,B4}}; see also \textrm{\cite{B5}} and
refs therein. In this regards, two remarkable approaches were given, first
by Creutz suggesting an extension of graphene dispersion relations by using
quaternions \textrm{\cite{B1,B4};} and subsequently by \emph{%
Bedaque-Bachoff-Tiburzi-WalkerLoud} (BBTW) \textrm{\cite{B2} proposing a 
\emph{4D}\ hyperdiamond lattice action with enough symmetries to exclude
fine tuning}. Apparently those two attempts look very close since both of
them extend \emph{2D} graphene to \emph{4D}; however they have basic
differences; some of them are discussed in \textrm{\cite{B2}}. The Creutz
model involves a two parameter lattice action that lives on a \emph{distorted%
} \emph{4D }lattice; and so looses the high discrete symmetry of the \emph{4D%
} hyperdiamond. The lattice action of BBTW model extends pristine \emph{2D}
graphene; it is built on perfect \emph{4D} hyperdiamond and has sufficient
discrete symmetries for a good continuum limit. Nevertheless, in both Creutz
and BBTW constructions, the distorted and perfect \emph{4D} hyperdiamonds
are thought of as made by the superposition of two sublattices $\mathcal{A}%
_{4}$ and $\mathcal{B}_{4}$ with massless left and right-handed fermions as
required by the no-go theorems for lattice chiral symmetry \textrm{\cite%
{C1,C2}}.\newline
Guided by the rich symmetries of the \emph{4D} hyperdiamond $\mathcal{H}_{4}$%
, we revisit in this paper, the BBTW model of ref \textrm{\cite{B2}} and its
higher dimension extensions given in \textrm{\cite{B5}} by using the hidden $%
SU\left( 5\right) $ [ resp. $SU\left( d+1\right) $] symmetry of $\mathcal{H}%
_{4}$ [ resp. $\mathcal{H}_{d+1}$ ] and its reciprocal lattice $\mathcal{H}%
_{4}^{\ast}$ [ resp. $\mathcal{H}_{d+1}^{\ast}$ ]. Focussing on \emph{4D}
lattice QCD, we first review the link between BBTW construction and $%
SU\left( 5\right) $. Then we refine the hyperdiamond lattice action by using
the weight vectors $\mathbf{\lambda}_{1},$ $\mathbf{\lambda}_{2},$ $\mathbf{%
\lambda}_{3},$ $\mathbf{\lambda}_{4},$ $\mathbf{\lambda}_{5}$ of the \emph{5}%
-dimensional (fundamental) representation of $SU\left( 5\right) $ as well as
mimicking pristine \emph{2D} graphene which, in the language of groups,
corresponds precisely to $SU\left( 3\right) $. After that, we study
explicitly the solutions of the zeros of the Dirac operator by using the $%
SU\left( 5\right) $ simple roots $\mathbf{\alpha}_{1},$ $\mathbf{\alpha}_{2},
$ $\mathbf{\alpha }_{3},$ $\mathbf{\alpha}_{4}$ generating $\mathcal{H}_{4}$%
; and its fundamental weights $\omega_{1},$ $\omega_{2},$ $\omega_{3},$ $%
\omega_{4}$ generating the reciprocal lattice $\mathcal{H}_{4}^{\ast}$. We
also comment the differences with BBTW construction; and exhibit the link
between the Dirac operator, following from our approach, and the one
suggested by Creutz using quaternions.\newline
The presentation is as follows: In section 2, we review briefly the BBTW
parametrization of the real 4D hyperdiamond $\mathcal{H}_{4}$ and comment
some particular discrete symmetries. In section 3, we study the link \
between $\mathcal{H}_{4}$ and the SU$\left( 5\right) $ symmetry. It is shown
that $\mathcal{H}_{4}$ is precisely generated by the four simple roots $%
\mathbf{\alpha}_{1},$ $\mathbf{\alpha}_{2},$ $\mathbf{\alpha}_{3},$ $\mathbf{%
\alpha}_{4}$ of $SU\left( 5\right) $; and the reciprocal lattice $\mathcal{H}%
_{4}^{\ast}$ is generated by its four weight vectors $\mathbf{\omega}_{1},$ $%
\mathbf{\omega}_{2},$ $\mathbf{\omega}_{3},$ $\mathbf{\omega}_{4}$. In
section 4, we revisit the BBTW\ model on $\mathcal{H}_{4}$ given in \textrm{%
\cite{B2}} and propose a refined 4D lattice action mimicking perfectly 2D
graphene. In section 5, we study explicitly the zeros of the Dirac operator;
and in section 6 we re-derive the Bori\c{c}i-Creutz fermions. In last
section, we give a conclusion and make comments regarding other lattice
models.

\section{\emph{4D} hyperdiamond $\mathcal{H}_{4}$}

Seen that the \emph{4D} hyperdiamond $\mathcal{H}_{4}$ plays a central role
in both BBTW and Creutz lattice models \textrm{\cite{B1,B2}}, we start by
studying this \emph{4D} lattice by exhibiting explicitly its
crystallographic structure. In particular, we give the relative positions of
the $\emph{5}$ first and the \emph{20} second nearest neighbors and exhibit
some particular discrete symmetries of $\mathcal{H}_{4}$. \newline
This analysis, which is useful in studying the link between the lattices $%
\mathcal{H}_{4}$ and the $SU\left( 5\right) $ simple roots, is important in
our construction; it will be used in section 3 to build the reciprocal
lattice $\mathcal{H}_{4}^{\ast}$ and in section 5 to study the dispersion
energy relations as well as the zeros of the Dirac operator.

\subsection{BBTW parametrization of $\mathcal{H}_{4}$}

In order to apply graphene simulation methods to lattice QCD, BBTW
generalizes tight binding model of \emph{2D} graphene to the \emph{4D}
diamond $\mathcal{H}_{4}$ \textrm{\cite{B2,BB2}; see also \cite{BA,B4,BCA,BD}%
}. Like in the case of \emph{2D} honeycomb, this \emph{4D} lattice is
defined by two superposed sublattices $\mathcal{A}_{4}$ and $\mathcal{B}_{4}$
with the two following basic objects: \newline
First, sites in $\mathcal{A}_{4}$ and $\mathcal{B}_{4}$ (L-nodes and R-nodes
in the terminology of \textrm{\cite{B2}}) are parameterized by the typical 
\emph{4d}- vectors $\mathbf{r}_{\mathbf{n}}$ with $\mathbf{n}=\left( {\small %
n}_{1}{\small ,n}_{2}{\small ,n}_{3}{\small ,n}_{4}\right) $ and ${\small n}%
_{i}$'s arbitrary integers. These lattice vectors are expanded as follows 
\begin{equation}
\begin{tabular}{llll}
$\mathcal{A}_{4}$ & : & $\mathbf{r}_{\mathbf{n}}=n_{1}$ $\mathbf{a}_{1}+n_{2}
$ $\mathbf{a}_{2}+n_{3}$ $\mathbf{a}_{3}+n_{4}$ $\mathbf{a}_{4}$ & $,$ \\ 
$\mathcal{B}_{4}$ & : & $\mathbf{r}_{\mathbf{n}}^{\prime}=\mathbf{r}_{%
\mathbf{n}}+\mathbf{e}$ & ,%
\end{tabular}
\label{Z1}
\end{equation}
where $\mathbf{a}_{1},$ $\mathbf{a}_{2},$ $\mathbf{a}_{3},$ $\mathbf{a}_{4}$
are primitive vectors generating these sublattices; and $\mathbf{e}$ is a
shift vector as described in what follows.\newline
Second, the vector $\mathbf{e}$ is a global vector taking the same value $%
\forall$ $\mathbf{n}$; it is a shift vector giving the relative positions of
the $\mathcal{B}_{4}$ sites with respect to the $\mathcal{A}_{4}$ ones; i.e: 
$\mathbf{e=r}_{\mathbf{n}}^{\prime}-\mathbf{r}_{\mathbf{n}}$, $\forall$ $%
\mathbf{n}$. In ref.\textrm{\cite{B2},} the $\mathbf{a}_{l}$'s and the $%
\mathbf{e}$ have been chosen as given by the following \emph{4}- component
vectors 
\begin{equation}
\begin{tabular}{lllllllll}
$\mathbf{a}_{1}$ & $=\mathbf{e}_{1}-\mathbf{e}_{5}$ & , & $\mathbf{a}_{3}$ & 
$=\mathbf{e}_{3}-\mathbf{e}_{5}$ & , & $\mathbf{e}$ & $=\mathbf{e}_{5}$ & ,
\\ 
$\mathbf{a}_{2}$ & $=\mathbf{e}_{2}-\mathbf{e}_{5}$ & , & $\mathbf{a}_{4}$ & 
$=\mathbf{e}_{4}-\mathbf{e}_{5}$ &  &  &  & 
\end{tabular}
\label{22}
\end{equation}
with the representation%
\begin{equation}
\begin{tabular}{lll}
$\mathbf{e}_{1}^{\mu}$ & $=\frac{1}{4}\left( +\sqrt{5},+\sqrt{5},+\sqrt {5}%
,+1\right) $ & , \\ 
&  &  \\ 
$\mathbf{e}_{2}^{\mu}$ & $=\frac{1}{4}\left( +\sqrt{5},-\sqrt{5},-\sqrt {5}%
,+1\right) $ & , \\ 
&  &  \\ 
$\mathbf{e}_{3}^{\mu}$ & $=\frac{1}{4}\left( -\sqrt{5},-\sqrt{5},+\sqrt {5}%
,+1\right) $ & , \\ 
&  &  \\ 
$\mathbf{e}_{4}^{\mu}$ & $=\frac{1}{4}\left( -\sqrt{5},+\sqrt{5},-\sqrt {5}%
,+1\right) $ & ,%
\end{tabular}
\label{Z2}
\end{equation}
and%
\begin{equation}
\begin{tabular}{lll}
$\mathbf{e}_{5}^{\mu}$ & $=-\mathbf{e}_{1}^{\mu}-\mathbf{e}_{2}^{\mu }-%
\mathbf{e}_{3}^{\mu}-\mathbf{e}_{4}^{\mu}$ & $=\left( \text{ \ }0,\text{ \ }%
0,\text{ \ }0,\text{ \ }-1\right) $.%
\end{tabular}
\label{ZZ}
\end{equation}
Notice also that the \emph{5} vectors $\mathbf{e}_{1},$ $\mathbf{e}_{2},$ $%
\mathbf{e}_{3},$ $\mathbf{e}_{4},$ $\mathbf{e}_{5}$ define the first nearest
neighbors to $\left( 0,0,0,0\right) $ and satisfy the constraint relations, 
\begin{equation}
\begin{tabular}{lll}
$\mathbf{e}_{i}.\mathbf{e}_{i}$ & $=\sum e_{i}^{\mu}.e_{i}^{\mu}=\sum
e_{i\mu }.e_{i}^{\mu}=1$ &  \\ 
&  &  \\ 
$\mathbf{e}_{i}.\mathbf{e}_{j}$ & $=\cos\vartheta_{ij}=-\frac{1}{4},\quad
i\neq j$ & ,%
\end{tabular}
\label{Z3}
\end{equation}
showing that the $\mathbf{e}_{i}$'s are distributed in a symmetric way since
all the angles $\vartheta_{ij}$ are equal to $\arccos\left( -\frac{1}{4}%
\right) $; see also figure (\ref{5NE}) for illustration. \newline
In the matrix representation (\ref{Z2}-\ref{ZZ}), the free four vectors $%
\mathbf{e}_{1},$ $\mathbf{e}_{2},$ $\mathbf{e}_{3},$ $\mathbf{e}_{4}$ are
permuted amongst each others by the typical unimodular matrices $\mathcal{O}%
_{_{\left[ ij\right] }}$ acting as 
\begin{equation}
\begin{tabular}{ll}
$\mathbf{e}_{i}^{\mu}=\dsum \limits_{\nu=1}^{4}\left( \mathcal{O}_{_{\left[
ji\right] }}\right) _{\nu}^{\mu}\mathbf{e}_{j}^{\nu},$ & $i,j=1,2,3,4,$%
\end{tabular}%
\end{equation}
with,%
\begin{equation}
\begin{tabular}{ll}
$\mathcal{O}_{_{\left[ 21\right] }}=\left( 
\begin{array}{cccc}
1 & 0 & 0 & 0 \\ 
0 & -1 & 0 & 0 \\ 
0 & 0 & -1 & 0 \\ 
0 & 0 & 0 & 1%
\end{array}
\right) $, & $\mathcal{O}_{_{\left[ 32\right] }}=\left( 
\begin{array}{cccc}
-1 & 0 & 0 & 0 \\ 
0 & 1 & 0 & 0 \\ 
0 & 0 & -1 & 0 \\ 
0 & 0 & 0 & 1%
\end{array}
\right) .$%
\end{tabular}%
\end{equation}
These transformations leave invariant the vector $\mathbf{e}_{5}=-\left( 
\mathbf{e}_{1}+\mathbf{e}_{2}+\mathbf{e}_{3}+\mathbf{e}_{4}\right) $; they
are sub-symmetries of the permutation group generated by permutations of the
five $\mathbf{e}_{i}$'s. We also have 
\begin{equation}
\begin{tabular}{llllll}
$\mathcal{O}_{_{\left[ 21\right] }}$ & $=\mathcal{O}_{_{\left[ 43\right] }}$
& , & $\mathcal{O}_{_{\left[ 31\right] }}$ & $=\mathcal{O}_{_{\left[ 32%
\right] }}\mathcal{O}_{_{\left[ 21\right] }}$ &  \\ 
$\mathcal{O}_{_{\left[ 32\right] }}$ & $=\mathcal{O}_{_{\left[ 14\right] }}$
& , & $\mathcal{O}_{_{\left[ 41\right] }}$ & $=\mathcal{O}_{_{\left[ 43%
\right] }}\mathcal{O}_{_{\left[ 31\right] }}$ & 
\end{tabular}%
\end{equation}
together with other similar relations.

\subsection{Some specific properties}

From the figure (\ref{5NE}) representing the first nearest neighbors in the
4D hyperdiamond and their analog in 2D graphene, we learn that each $%
\mathcal{A}_{4}$- type node at $\mathbf{r}_{\mathbf{n}}$, with some attached
wave function $A_{\mathbf{r}_{\mathbf{n}}}$, has the following closed
neighbors: 
\begin{figure}[ptbh]
\begin{center}
\hspace{0cm} \includegraphics[width=12cm]{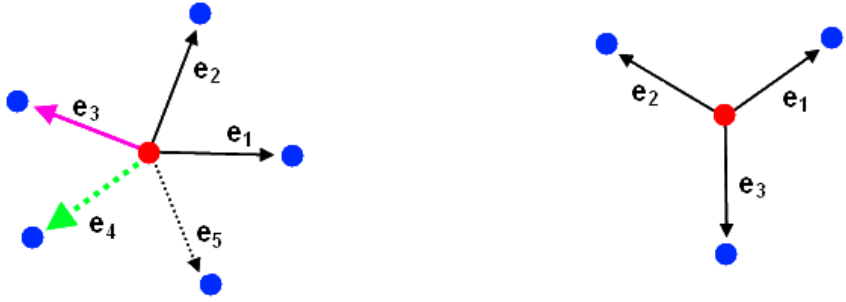}
\end{center}
\par
\vspace{-0.5 cm}
\caption{On left the \emph{5} first nearest neighbors in the pristine \emph{%
4D} hyperdiamond with the properties $\left\Vert \mathbf{e}_{i}\right\Vert
=1 $ and $\mathbf{e}_{1}+\mathbf{e}_{2}+\mathbf{e}_{3}+\mathbf{e}_{4}+%
\mathbf{e}_{5}=0$. On right, the \emph{3} first nearest in pristine \emph{2D}
graphene with $\left\Vert \mathbf{e}_{i}\right\Vert =1$ and $\mathbf{e}_{1}+%
\mathbf{e}_{2}+\mathbf{e}_{3}=0$. }
\label{5NE}
\end{figure}
\emph{5} first nearest neighbors belonging to $\mathcal{B}_{4}$ with wave
functions $B_{\mathbf{r}_{\mathbf{n}}+d\mathbf{e}_{i}}$; and \emph{20}
second nearest neighbors belonging to the same $\mathcal{A}_{4}$ with the
wave functions $A_{\mathbf{r}_{\mathbf{n}}+d\left( \mathbf{e}_{i}-\mathbf{e}%
_{j}\right) }$. The first nearest neighbors are given by:%
\begin{equation}
\begin{tabular}{llll}
\textit{lattice position} &  & \textit{attached wave} &  \\ 
$\ \ \ \mathbf{r}_{\mathbf{n}}+d\mathbf{e}_{1}$ & $\longleftrightarrow$ & $\
\ \ B_{\mathbf{r}_{\mathbf{n}}+d\mathbf{e}_{1}}$ &  \\ 
$\ \ \ \mathbf{r}_{\mathbf{n}}+d\mathbf{e}_{2}$ & $\longleftrightarrow$ & $\
\ \ B_{\mathbf{r}_{\mathbf{n}}+d\mathbf{e}_{2}}$ &  \\ 
$\ \ \ \mathbf{r}_{\mathbf{n}}+d\mathbf{e}_{3}$ & $\longleftrightarrow$ & $\
\ \ B_{\mathbf{r}_{\mathbf{n}}+d\mathbf{e}_{3}}$ &  \\ 
$\ \ \ \mathbf{r}_{\mathbf{n}}+d\mathbf{e}_{4}$ & $\longleftrightarrow$ & $\
\ \ B_{\mathbf{r}_{\mathbf{n}}+d\mathbf{e}_{4}}$ &  \\ 
$\ \ \ \mathbf{r}_{\mathbf{n}}+d\mathbf{e}_{5}$ & $\longleftrightarrow$ & $\
\ \ B_{\mathbf{r}_{\mathbf{n}}+d\mathbf{e}_{5}}$ & 
\end{tabular}%
\end{equation}
Using this configuration, the typical tight binding hamiltonian describing
the couplings between the first nearest neighbors reads as 
\begin{equation}
\begin{tabular}{ll}
$-t\dsum \limits_{\mathbf{r}_{\mathbf{n}}}\dsum \limits_{i=1}^{5}A_{\mathbf{r%
}_{\mathbf{n}}}B_{\mathbf{r}_{\mathbf{n}}+d\mathbf{e}_{i}}^{+}+hc$ & .%
\end{tabular}%
\end{equation}
where $t$ is the hop energy and where $d$ is the lattice parameter. Notice
that in the case where the wave functions at $\mathbf{r}_{\mathbf{n}}$ and $%
\mathbf{r}_{\mathbf{n}}+d\mathbf{e}_{i}$ are rather given by two component
Weyl spinors 
\begin{equation}
\begin{tabular}{llll}
$A_{\mathbf{r}_{\mathbf{n}}}^{a}=\left( 
\begin{array}{c}
A_{\mathbf{r}_{\mathbf{n}}}^{1} \\ 
A_{\mathbf{r}_{\mathbf{n}}}^{2}%
\end{array}
\right) $ & , & $\bar{B}_{\mathbf{r}_{\mathbf{n}}+d\mathbf{e}_{i}}^{\dot{a}%
}=\left( 
\begin{array}{c}
\bar{B}_{\mathbf{r}_{\mathbf{n}}+d\mathbf{e}_{i}}^{\dot{1}} \\ 
\bar{B}_{\mathbf{r}_{\mathbf{n}}+d\mathbf{e}_{i}}^{\dot{2}}%
\end{array}
\right) $ & ,%
\end{tabular}%
\end{equation}
together with their adjoints $\bar{A}_{\mathbf{r}_{\mathbf{n}}}^{\dot{a}}$
and $\bar{B}_{\mathbf{r}_{\mathbf{n}}+d\mathbf{e}_{i}}^{a}$, as in the
example of 4D lattice QCD to be described in section 4, the corresponding
tight binding model would be, 
\begin{equation}
\begin{tabular}{ll}
$-t\dsum \limits_{\mathbf{r}_{\mathbf{n}}}\dsum \limits_{i=1}^{5}\left[
\dsum \limits_{\mu=1}^{4}\mathbf{e}_{i}^{\mu}\left( A_{\mathbf{r}_{\mathbf{n}%
}}^{a}\mathrm{\sigma }_{a\dot{a}}^{\mu}\bar{B}_{\mathbf{r}_{\mathbf{n}}+d%
\mathbf{e}_{i}}^{\dot{a}}\right) \right] +hc$ & .%
\end{tabular}%
\end{equation}
where the $\mathbf{e}_{i}^{\mu}$'s are as in (\ref{Z2}) and where the
coefficients $\mathrm{\sigma}_{a\dot{a}}^{\mu}$ will be specified later on.
Notice moreover that the term $\sum_{i=1}^{5}\mathbf{e}_{i}^{\mu}\left( A_{%
\mathbf{r}_{\mathbf{n}}}^{a}\mathrm{\sigma}_{a\dot{a}}^{\mu}\bar {B}_{%
\mathbf{r}_{\mathbf{n}}}^{\dot{a}}\right) $ vanishes identically due to $%
\sum_{i=1}^{5}\mathbf{e}_{i}^{\mu}=0.$ The \emph{20} second nearest
neighbors read as%
\begin{equation}
\begin{tabular}{llll}
${\small r}_{\mathbf{n}}\pm d\left( \mathbf{e}_{1}-\mathbf{e}_{2}\right) 
{\small ,}$ & ${\small r}_{\mathbf{n}}\pm d\left( \mathbf{e}_{1}-\mathbf{e}%
_{3}\right) {\small ,}$ & ${\small r}_{\mathbf{n}}\pm d\left( \mathbf{e}_{1}-%
\mathbf{e}_{4}\right) {\small ,}$ &  \\ 
${\small r}_{\mathbf{n}}\pm d\left( \mathbf{e}_{1}-\mathbf{e}_{5}\right) 
{\small ,}$ & ${\small r}_{\mathbf{n}}\pm d\left( \mathbf{e}_{2}-\mathbf{e}%
_{3}\right) {\small ,}$ & ${\small r}_{\mathbf{n}}\pm d\left( \mathbf{e}_{2}-%
\mathbf{e}_{4}\right) {\small ,}$ &  \\ 
${\small r}_{\mathbf{n}}\pm d\left( \mathbf{e}_{2}-\mathbf{e}_{5}\right) 
{\small ,}$ & ${\small r}_{\mathbf{n}}\pm d\left( \mathbf{e}_{3}-\mathbf{e}%
_{4}\right) {\small ,}$ & ${\small r}_{\mathbf{n}}\pm d\left( \mathbf{e}_{3}-%
\mathbf{e}_{5}\right) {\small ,}$ &  \\ 
${\small r}_{\mathbf{n}}\pm d\left( \mathbf{e}_{4}-\mathbf{e}_{5}\right) 
{\small .}$ &  &  & 
\end{tabular}
\label{Z4}
\end{equation}
At this order, the standard tight binding hamiltonian reads as follows 
\begin{equation}
\begin{tabular}{ll}
$-t^{\prime}\dsum \limits_{\mathbf{r}_{\mathbf{n}}}\dsum
\limits_{i,j=1}^{5}\left( A_{\mathbf{r}_{\mathbf{n}}}A_{\mathbf{r}_{\mathbf{n%
}}+d\left( \mathbf{e}_{i}-\mathbf{e}_{j}\right) }^{+}+B_{\mathbf{r}_{\mathbf{%
n}}}B_{\mathbf{r}_{\mathbf{n}}+d\left( \mathbf{e}_{i}-\mathbf{e}_{j}\right)
}^{+}\right) +hc$ & .%
\end{tabular}%
\end{equation}
and in the case of Weyl spinors, we have%
\begin{equation}
-t^{\prime}\dsum \limits_{\mathbf{r}_{\mathbf{n}}}\dsum \limits_{i,j=1}^{5}%
\left[ \dsum \limits_{\mu=1}^{4}\mathbf{e}_{i}^{\mu}\left( A_{\mathbf{r}_{%
\mathbf{n}}}^{a}\mathrm{\sigma }_{a\dot{a}}^{\mu}\bar{A}_{\mathbf{r}_{%
\mathbf{n}}+d\left( \mathbf{e}_{i}-\mathbf{e}_{j}\right) }^{\dot{a}}+B_{%
\mathbf{r}_{\mathbf{n}}}^{a}\mathrm{\sigma}_{a\dot{a}}^{\mu}\bar{B}_{\mathbf{%
r}_{\mathbf{n}}+d\left( \mathbf{e}_{i}-\mathbf{e}_{j}\right) }^{\dot{a}%
}\right) \right] +hc
\end{equation}
In what follows, we show that the \emph{5} vectors $\mathbf{e}_{1},$ $%
\mathbf{e}_{2},$ $\mathbf{e}_{3},$ $\mathbf{e}_{4}$ $\mathbf{e}_{5}$ are, up
to a normalization factor namely $\frac{\sqrt{5}}{2}$, precisely the weight
vectors $\mathbf{\lambda}_{0},$ $\mathbf{\lambda}_{1},$ $\mathbf{\lambda}%
_{2},$ $\mathbf{\lambda}_{3},$ $\mathbf{\lambda}_{4}$ of the \emph{5}%
-dimensional representation of $SU\left( 5\right) $; and the \emph{20}
vectors $\left( \mathbf{e}_{i}-\mathbf{e}_{j}\right) $ are, up to a scale
factor $\frac{\sqrt{5}}{2}$, their roots $\beta_{ij}=\left( \mathbf{\lambda }%
_{i}-\mathbf{\lambda}_{j}\right) $. We show as well that the particular
property $\mathbf{e}_{i}.\mathbf{e}_{j}=-\frac{1}{4}$, which is constant $%
\forall$ $\mathbf{e}_{i}$, $\forall$ $\mathbf{e}_{j}$, has a natural
interpretation in terms of the Cartan matrix of $SU\left( 5\right) $.

\section{Link with $SU\left( 5\right) $ symmetry}

For later use, we exhibit here the hidden $SU\left( 5\right) $ symmetry of
the \emph{4D} hyperdiamond; we show that $\mathcal{H}_{4}$ considered above
is precisely the lattice $\mathcal{L}_{su\left( 5\right) }$ studied in 
\textrm{\cite{D1}}. More concretely, we show the three following:\newline
First, the \emph{5} bond vectors $\mathbf{e}_{1},$ $\mathbf{e}_{2},$ $%
\mathbf{e}_{3},$ $\mathbf{e}_{4},$ $\mathbf{e}_{5}$ ( first nearest
neighbors) are given by the \emph{5} weight vectors $\mathbf{\lambda}_{1},$ $%
\mathbf{\lambda}_{2},$ $\mathbf{\lambda}_{3},$ $\mathbf{\lambda}_{4},$ $%
\mathbf{\lambda}_{5}$ ( below, we set $\mathbf{\lambda}_{5}\equiv \mathbf{%
\lambda}_{0}$) of the 5-dimensional (fundamental) representation of $%
SU\left( 5\right) $ which also satisfy%
\begin{equation}
\mathbf{\lambda}_{0}+\mathbf{\lambda}_{1}+\mathbf{\lambda}_{2}+\mathbf{%
\lambda }_{3}+\mathbf{\lambda}_{4}=0   \label{LA}
\end{equation}
We will show later that $\mathbf{e}_{i}=\frac{\sqrt{5}}{2}\mathbf{\lambda}%
_{i}$ with $\mathbf{\lambda}_{i}.\mathbf{\lambda}_{i}=\frac{4}{5}.$\newline
Second, the \emph{4} primitive ones $\mathbf{a}_{1},$ $\mathbf{a}_{2},$ $%
\mathbf{a}_{3},$ $\mathbf{a}_{4}$ used in generating $\mathcal{H}_{4}$ are
particular linear combinations of the \emph{4} simple roots $\mathbf{\alpha }%
_{1},$ $\mathbf{\alpha}_{2},$ $\mathbf{\alpha}_{3},$ $\mathbf{\alpha}_{4}$
of $SU\left( 5\right) $; see eq(\ref{AA}) for the explicit relations. Recall
that the $SU\left( 5\right) $ symmetry has \emph{20} roots as given below, 
\begin{equation}
\begin{tabular}{llll}
$\pm\mathbf{\alpha}_{1},$ & $\pm\left( \mathbf{\alpha}_{1}+\mathbf{\alpha }%
_{2}\right) ,$ & $\pm\left( \mathbf{\alpha}_{1}+\mathbf{\alpha}_{2}+\mathbf{%
\alpha}_{3}\right) ,$ & $\pm\left( \mathbf{\alpha}_{1}+\mathbf{\alpha}_{2}+%
\mathbf{\alpha}_{3}+\mathbf{\alpha}_{4}\right) $ \\ 
$\pm\mathbf{\alpha}_{2},$ & $\pm\left( \mathbf{\alpha}_{2}+\mathbf{\alpha }%
_{3}\right) ,$ & $\pm\left( \mathbf{\alpha}_{2}+\mathbf{\alpha}_{3}+\mathbf{%
\alpha}_{4}\right) ,$ &  \\ 
$\pm\mathbf{\alpha}_{3},$ & $\pm\left( \mathbf{\alpha}_{3}+\mathbf{\alpha }%
_{4}\right) ,$ &  &  \\ 
$\pm\mathbf{\alpha}_{3}$ &  &  & 
\end{tabular}
\label{RO}
\end{equation}
These vectors have all of them the same length $\mathbf{\alpha}^{2}=2$; and
so they generate the relative lattice positions of the second nearest
neighbors in the \emph{4D} hyperdiamond.\newline
Third, the $SU\left( 5\right) $ has also discrete symmetries given by the so
called Weyl group transformations generated by the $\sigma_{\mathbf{\alpha}}$%
's acting on generic roots $\mathbf{\beta}$ of SU$\left( 5\right) $ as
follows,%
\begin{equation}
\begin{tabular}{llll}
$\sigma_{\mathbf{\alpha}}\left( \mathbf{\beta}\right) $ & $=$ & $\mathbf{%
\beta}-2\frac{\mathbf{\alpha.\beta}}{\mathbf{\alpha}^{2}}\mathbf{\beta=\beta}%
-\left( \alpha.\beta\right) $ $\mathbf{\beta}$ & .%
\end{tabular}%
\end{equation}
These discrete transformations permute the roots (\ref{RO}) amongst
themselves and are isomorphic to $\mathcal{S}_{5}$ permutation group
transformations. For instance, we have $\sigma_{\mathbf{\alpha}_{1}}\left( 
\mathbf{\alpha}_{1}\right) =-\mathbf{\alpha}_{1}$ and $\sigma_{\mathbf{\alpha%
}_{1}}\left( \mathbf{\alpha}_{2}\right) =\mathbf{\alpha}_{1}+\mathbf{\alpha}%
_{2}$.

\subsection{Exhibiting the link $\mathcal{H}_{4}/SU\left( 5\right) $}

To exhibit explicitly the link between pristine lattice $\mathcal{H}_{4}$
and the simple roots and the basic weight vectors of $SU\left( 5\right) $,
we start by recalling some of its features; in particular the following
useful ingredients: $SU\left( 5\right) $ is a \emph{24} dimensional symmetry
group; it has rank \emph{4}; that is \emph{4} simple roots $\mathbf{\alpha}%
_{1},$ $\mathbf{\alpha}_{2},$ $\mathbf{\alpha}_{3},$ $\mathbf{\alpha}_{4}$;
it has \emph{20} roots\ $\pm\mathbf{\beta}_{ij}$ given by eq(\ref{RO}). The
simple roots $\mathbf{\alpha}_{1},$ $\mathbf{\alpha}_{2},$ $\mathbf{\alpha}%
_{3},$ $\mathbf{\alpha}_{4}$ capture most of the algebraic properties of SU$%
\left( 5\right) $; and as a consequence, those of the \emph{4D} hyperdiamond
crystal; in particular they generate the \emph{20} roots\ $\pm\mathbf{\beta }%
_{ij}$ as shown on (\ref{RO}) and they have a symmetric intersection matrix $%
\mathbf{K}_{ij}=\mathbf{\alpha}_{i}$.$\mathbf{\alpha}_{j}$ with inverse $%
\mathbf{K}_{ij}^{-1}$ given by, 
\begin{equation}
\begin{tabular}{llll}
$\mathbf{K}_{ij}=\left( 
\begin{array}{cccc}
2 & -1 & 0 & 0 \\ 
-1 & 2 & -1 & 0 \\ 
0 & -1 & 2 & -1 \\ 
0 & 0 & -1 & 2%
\end{array}
\right) $ & , & $\mathbf{K}_{ij}^{-1}=\frac{1}{5}\left( 
\begin{array}{cccc}
4 & 3 & 2 & 1 \\ 
3 & 6 & 4 & 2 \\ 
2 & 4 & 6 & 3 \\ 
1 & 2 & 3 & 4%
\end{array}
\right) $ & 
\end{tabular}
\label{Z5}
\end{equation}
that encode the algebraic data of the underlying Lie algebra of the $%
SU\left( 5\right) $ symmetry. These simple roots define as well the \emph{4}
fundamental weights $\mathbf{\omega}_{1},$ $\mathbf{\omega}_{2},$ $\mathbf{%
\omega}_{3},$ $\mathbf{\omega}_{4}$ through the following duality relation%
\begin{equation}
\begin{tabular}{llll}
$\mathbf{\omega}_{i}.\mathbf{\alpha}_{j}$ & $=$ & $\delta_{ij},$ & $%
i,j=1,...,4$.%
\end{tabular}
\label{wa}
\end{equation}
These fundamental weights are important for us; first because they allow to
build the reciprocal \emph{4D} hyperdiamond $\mathcal{H}_{4}^{\ast}$ and
second can be use used to expand any wave vector in $\mathcal{H}_{4}^{\ast}$
as follows%
\begin{equation}
\mathbf{k}=k_{1}\mathbf{\omega}_{1}+k_{2}\mathbf{\omega}_{2}+k_{3}\mathbf{%
\omega}_{3}+k_{4}\mathbf{\omega}_{4}.   \label{ka}
\end{equation}
From this expansion we read the relations $k_{i}=\mathbf{k}.\mathbf{\alpha }%
_{i}$ showing that the $k_{i}$'s are precisely the wave vector components
propagating along the $\mathbf{\alpha}_{i}$-directions; thanks to eqs(\ref%
{wa}).

\subsection{Other useful relations}

Using the matrices $\mathbf{K}_{ij}$ and $\mathbf{K}_{ij}^{-1}$, one can
express the simple roots $\mathbf{\alpha}_{i}$ in terms of the fundamental
weight vectors $\mathbf{\omega}_{i}$; and inversely the $\mathbf{\omega}_{i}$%
's as linear combinations of the simple roots as given below,%
\begin{equation}
\begin{tabular}{ll}
$\mathbf{\omega}_{1}=\frac{4}{5}\mathbf{\alpha}_{1}+\frac{3}{5}\mathbf{%
\alpha }_{2}+\frac{2}{5}\mathbf{\alpha}_{3}+\frac{1}{5}\mathbf{\alpha}_{4}\ $
&  \\ 
$\mathbf{\omega}_{2}=\frac{3}{5}\mathbf{\alpha}_{1}+\frac{6}{5}\mathbf{%
\alpha }_{2}+\frac{4}{5}\mathbf{\alpha}_{3}+\frac{2}{5}\mathbf{\alpha}_{4}\ $
&  \\ 
$\mathbf{\omega}_{3}=\frac{2}{5}\mathbf{\alpha}_{1}+\frac{4}{5}\mathbf{%
\alpha }_{2}+\frac{6}{5}\mathbf{\alpha}_{3}+\frac{3}{5}\mathbf{\alpha}_{4}\ $
&  \\ 
$\mathbf{\omega}_{4}=\frac{1}{5}\mathbf{\alpha}_{1}+\frac{2}{5}\mathbf{%
\alpha }_{2}+\frac{3}{5}\mathbf{\alpha}_{3}+\frac{4}{5}\mathbf{\alpha}_{4}\ $
& 
\end{tabular}
\label{om}
\end{equation}
Using these relations, it is not difficult to check that they satisfy (\ref%
{wa}); for instance we have $\mathbf{\omega}_{1}.\mathbf{\alpha}_{1}$ $=$ $%
\frac{8}{5}-\frac{3}{5}=1$ and $\mathbf{\omega}_{1}.\mathbf{\alpha}_{2}$ $= $
$-\frac{4}{5}+\frac{6}{5}-\frac{2}{5}$ $=0$; and similarly for the others $%
\mathbf{\omega}_{2},$ $\mathbf{\omega}_{3},$ $\mathbf{\omega}_{4}$ and the
intersections $\mathbf{\omega}_{i}.\mathbf{\alpha}_{j}$. Notice moreover
that the fundamental weight vector $\mathbf{\omega}_{1}$ defines a highest
weight representation of $SU\left( 5\right) $ of dimension \emph{5} with
weight vectors $\mathbf{\lambda}_{0},$ $\mathbf{\lambda}_{1},$ $\mathbf{%
\lambda}_{2},$ $\mathbf{\lambda}_{3},$ $\mathbf{\lambda}_{4}$ related to $%
\mathbf{\omega}_{1}$ as follows 
\begin{equation}
\begin{tabular}{lllll}
$\mathbf{\lambda}_{0}$ & $=$ & $\mathbf{\omega}_{1}$ & , &  \\ 
$\mathbf{\lambda}_{1}$ & $=$ & $\mathbf{\omega}_{1}-\mathbf{\alpha}_{1}$ & ,
&  \\ 
$\mathbf{\lambda}_{2}$ & $=$ & $\mathbf{\omega}_{1}-\mathbf{\alpha}_{1}-%
\mathbf{\alpha}_{2}$ & , &  \\ 
$\mathbf{\lambda}_{3}$ & $=$ & $\mathbf{\omega}_{1}-\mathbf{\alpha}_{1}-%
\mathbf{\alpha}_{2}-\mathbf{\alpha}_{3}$ & , &  \\ 
$\mathbf{\lambda}_{4}$ & $=$ & $\mathbf{\omega}_{1}-\mathbf{\alpha}_{1}-%
\mathbf{\alpha}_{2}-\mathbf{\alpha}_{3}-\mathbf{\alpha}_{4}\ $ & . & 
\end{tabular}
\label{lam}
\end{equation}
By using (\ref{om}), one may also express these vectors weights in terms of
the $\mathbf{\omega}_{i}$'s as follows%
\begin{equation}
\begin{tabular}{lllll}
$\mathbf{\lambda}_{0}$ & $=$ & $\mathbf{\omega}_{1}$ & , &  \\ 
$\mathbf{\lambda}_{1}$ & $=$ & $\mathbf{\omega}_{2}-\mathbf{\omega}_{1}$ & ,
&  \\ 
$\mathbf{\lambda}_{2}$ & $=$ & $\mathbf{\omega}_{3}-\mathbf{\omega}_{2}$ & ,
&  \\ 
$\mathbf{\lambda}_{3}$ & $=$ & $\mathbf{\omega}_{4}-\mathbf{\omega}_{3}$ & ,
&  \\ 
$\mathbf{\lambda}_{4}$ & $=$ & $-\mathbf{\omega}_{4}\ $ & . & 
\end{tabular}
\label{lw}
\end{equation}
Furthermore, substituting $\mathbf{\omega}_{1}$ by its expression (\ref{om}%
), we get the following values of the $\mathbf{\lambda}_{i}$'s in terms of
the simple roots 
\begin{equation}
\begin{tabular}{ll}
$\mathbf{\lambda}_{0}=+\frac{4}{5}\mathbf{\alpha}_{1}+\frac{3}{5}\mathbf{%
\alpha}_{2}+\frac{2}{5}\mathbf{\alpha}_{3}+\frac{1}{5}\mathbf{\alpha }_{4}\ $
& , \\ 
$\mathbf{\lambda}_{1}=-\frac{1}{5}\mathbf{\alpha}_{1}+\frac{3}{5}\mathbf{%
\alpha}_{2}+\frac{2}{5}\mathbf{\alpha}_{3}+\frac{1}{5}\mathbf{\alpha }_{4}\ $
& , \\ 
$\mathbf{\lambda}_{2}=-\frac{1}{5}\mathbf{\alpha}_{1}-\frac{2}{5}\mathbf{%
\alpha}_{2}+\frac{2}{5}\mathbf{\alpha}_{3}+\frac{1}{5}\mathbf{\alpha }_{4}\ $
& , \\ 
$\mathbf{\lambda}_{3}=-\frac{1}{5}\mathbf{\alpha}_{1}-\frac{2}{5}\mathbf{%
\alpha}_{2}-\frac{3}{5}\mathbf{\alpha}_{3}+\frac{1}{5}\mathbf{\alpha }_{4}\ $
& , \\ 
$\mathbf{\lambda}_{4}=-\frac{1}{5}\mathbf{\alpha}_{1}-\frac{2}{5}\mathbf{%
\alpha}_{2}-\frac{3}{5}\mathbf{\alpha}_{3}-\frac{4}{5}\mathbf{\alpha }_{4}\ $
& .%
\end{tabular}
\label{Z6}
\end{equation}
These weight vectors satisfy remarkable properties that will be used later
on; in particular the three following: First, these $\mathbf{\lambda}_{i}$'s
obey the constraint relation $\sum_{i=0}^{4}\mathbf{\lambda}_{i}=0$ which
agrees with (\ref{LA}) and which should be compared with the identity $%
\mathbf{e}_{1}^{\mu}+\mathbf{e}_{2}^{\mu}+\mathbf{e}_{3}^{\mu}+\mathbf{e}%
_{4}^{\mu }+\mathbf{e}_{4}^{\mu}=0$. Second, they have the intersection
matrix%
\begin{equation}
\begin{tabular}{llllll}
$\mathbf{\lambda}_{i}.\mathbf{\lambda}_{i}=\frac{4}{5}$ & , & $\mathbf{%
\lambda }_{i}.\mathbf{\lambda}_{j}=-\frac{1}{5}$ & , & $\cos\vartheta_{ij}=%
\frac{\mathbf{\lambda}_{i}.\mathbf{\lambda}_{j}}{\left\vert \mathbf{\lambda }%
_{i}\right\vert \left\vert \mathbf{\lambda}_{j}\right\vert }=-\frac{1}{4}$ & 
,%
\end{tabular}
\label{Z7}
\end{equation}
leading to eq(\ref{Z3}). The third point concerns the zeros of the Dirac
operator; see eq(\ref{D1}) to fix the ideas. They are given by solving the
following constraint relations 
\begin{equation}
\begin{tabular}{ll}
$e^{i\frac{d\sqrt{5}}{2}p_{0}}=e^{i\frac{d\sqrt{5}}{2}p_{1}}=e^{i\frac {d%
\sqrt{5}}{2}p_{2}}=e^{i\frac{d\sqrt{5}}{2}p_{3}}=e^{i\frac{d\sqrt{5}}{2}%
p_{4}}$ & $=e^{i\mathbf{\varphi}},$%
\end{tabular}
\label{cst}
\end{equation}
where we have set%
\begin{equation}
\begin{tabular}{llllll}
$p_{0}=\mathbf{k.\lambda}_{0}$ & , & $p_{1}=\mathbf{k.\lambda}_{1}$ & , & $%
p_{2}=\mathbf{k.\lambda}_{2}$ &  \\ 
$p_{3}=\mathbf{k.\lambda}_{3}$ & , & $p_{4}=\mathbf{k.\lambda}_{4}$ &  &  & 
\end{tabular}
\label{pi}
\end{equation}
and where the phase $\varphi=\frac{2\pi N}{5}$, with $N$ an integer. The
values of this phase are due to equiprobability in hops from a generic site
at $\mathbf{r}$ to the \emph{5} first nearest neighbors at $\mathbf{r+}\frac{%
d\sqrt{5}}{2}\mathbf{\lambda}_{i}$. This equiprobability requires%
\begin{equation}
\dprod \limits_{l=0}^{5}e^{i\frac{d\sqrt{5}}{2}p_{l}}=1=e^{5i\mathbf{\varphi}%
}.   \label{ph}
\end{equation}
Solutions of the constraint eqs(\ref{cst}) are then given by 
\begin{equation}
\begin{tabular}{llll}
$p_{i}=\frac{4\pi N}{5d\sqrt{5}}$ & , & $i=0,1,2,3,4$ & .%
\end{tabular}
\label{S2}
\end{equation}
Notice moreover the two useful features: First, eqs(\ref{pi}) imply in turn
that the wave vector $\mathbf{k}$ may be also written as 
\begin{equation}
\mathbf{k}=p_{0}\mathbf{\lambda}_{0}+p_{1}\mathbf{\lambda}_{1}+p_{2}\mathbf{%
\lambda}_{2}+p_{3}\mathbf{\lambda}_{3}+p_{4}\mathbf{\lambda}_{4}   \label{X}
\end{equation}
Multiplying both sides of this relation by $\mathbf{\lambda}_{i}$ and using (%
\ref{Z7}), we find $\mathbf{k.\lambda}_{i}$ $=$ $p_{i}-$ $\frac{1}{5}\left(
p_{0}+p_{1}+p_{2}+p_{3}+p_{4}\right) $ $=$ $p_{i}$; thanks to the identity $%
\sum_{i}p_{i}=0$ following from $\sum_{i}\mathbf{\lambda}_{i}=0$. Second,
expressing this vector $\mathbf{k}$ in terms of the basis $\mathbf{\omega}%
_{1},$ $\mathbf{\omega}_{2},$ $\mathbf{\omega}_{3},$ $\mathbf{\omega}_{4}$
of the reciprocal lattice; then using eqs(\ref{lw}) giving the $\mathbf{%
\lambda }_{l}$'s in terms of the $\mathbf{\omega}_{l}$'s, we get%
\begin{equation}
\begin{tabular}{lll}
$\mathbf{k}$ & $=\left( p_{0}-p_{1}\right) \mathbf{\omega}_{1}+\left(
p_{1}-p_{2}\right) \mathbf{\omega}_{2}+\left( p_{2}-p_{3}\right) \mathbf{%
\omega}_{3}+\left( p_{3}-p_{4}\right) \mathbf{\omega}_{4}$ & .%
\end{tabular}
\label{Y}
\end{equation}
Putting back (\ref{S2}), we find that the zeros of the Dirac operator are
precisely located at the sites of the reciprocal lattice $\mathcal{H}%
_{4}^{\ast}$.

\subsection{Link with BBTW parametrization of\emph{\ }$\mathcal{H}_{\emph{4}%
} $}

From eq(\ref{lam}), we can also determine the expression of the simple roots 
$\mathbf{\alpha}_{i}$'s in terms of the weight vectors $\mathbf{\lambda}_{i}$%
's; we have:%
\begin{equation}
\begin{tabular}{llll}
$\mathbf{\alpha}_{1}=\mathbf{\lambda}_{0}-\mathbf{\lambda}_{1}$ & , & $%
\mathbf{\alpha}_{3}=\mathbf{\lambda}_{2}-\mathbf{\lambda}_{3}$ & , \\ 
$\mathbf{\alpha}_{2}=\mathbf{\lambda}_{1}-\mathbf{\lambda}_{2}$ & , & $%
\mathbf{\alpha}_{4}=\mathbf{\lambda}_{3}-\mathbf{\lambda}_{4}$ & .%
\end{tabular}
\label{Z8}
\end{equation}
By comparing these equations with eq(\ref{Z2}-\ref{Z3}), we obtain the
relation between the $\mathbf{e}_{i}$'s used in \textrm{\cite{B2}} and the
weight vectors of the fundamental representation of $SU\left( 5\right) $; 
\begin{equation}
\begin{tabular}{llll}
$\mathbf{e}_{i}=\frac{\sqrt{5}}{2}\mathbf{\lambda}_{i}$ & , & $\mathbf{%
\lambda }_{i}=\frac{2\sqrt{5}}{5}\mathbf{e}_{i}$ & ,%
\end{tabular}
\label{Z9}
\end{equation}
Putting eqs(\ref{Z8},\ref{Z9}) back into (\ref{22}), we find that the \emph{4%
} primitive vectors $\mathbf{a}_{1}$, $\mathbf{a}_{2}$, $\mathbf{a}_{3}$, $%
\mathbf{a}_{4}$ generating the sublattice $\mathcal{A}_{4}$ (resp. $\mathcal{%
B}_{4}$) are nothing but linear combinations of the four simple roots of $%
SU\left( 5\right) $, 
\begin{equation}
\begin{tabular}{ll}
$\mathbf{a}_{1}=$ & $-\frac{\sqrt{5}}{2}\mathbf{\alpha}_{1}$ \\ 
$\mathbf{a}_{2}=$ & $-\frac{\sqrt{5}}{2}\left( \mathbf{\alpha}_{1}+\mathbf{%
\alpha}_{2}\right) $ \\ 
$\mathbf{a}_{3}=$ & $-\frac{\sqrt{5}}{2}\left( \mathbf{\alpha}_{1}+\mathbf{%
\alpha}_{2}+\mathbf{\alpha}_{3}\right) $ \\ 
$\mathbf{a}_{4}=$ & $-\frac{\sqrt{5}}{2}\left( \mathbf{\alpha}_{1}+\mathbf{%
\alpha}_{2}+\mathbf{\alpha}_{3}+\mathbf{\alpha}_{4}\right) $%
\end{tabular}
\label{AA}
\end{equation}
From these relations, we read the identities of \cite{B2}%
\begin{equation}
\begin{tabular}{lllll}
$\mathbf{a}_{i}.\mathbf{a}_{i}=\frac{10}{4}$ & , & $\mathbf{a}_{i}.\mathbf{a}%
_{j}=\frac{5}{4},$ & $i\neq j$ & .%
\end{tabular}
\label{Z10}
\end{equation}
These relations are just a property of Cartan matrix of $SU\left( 5\right) $.%
\newline
We end this section by giving the following summary:\newline
The \emph{4D} hyperdiamond $\mathcal{H}_{4}$ is made of two superposed
sublattices $\mathcal{A}_{4}$ and $\mathcal{B}_{4}$. These sublattices are
generated by the simple roots $\mathbf{\alpha}_{1},$ $\mathbf{\alpha}_{2},$ $%
\mathbf{\alpha }_{3},$ $\mathbf{\alpha}_{4}$ of $SU\left( 5\right) $. The
relative shift vector between $\mathcal{A}_{4}$ and $\mathcal{B}_{4}$ is a
weight vector of the \emph{5}-dimensional representation of $SU\left(
5\right) $. Each site in $\mathcal{H}_{4}$ has \emph{5} first nearest
neighbors forming a dimension \emph{5} representation of $SU\left( 5\right) $%
; and \emph{20} second nearest ones; forming together with the "\emph{4}
zero roots", the adjoint representation of $SU\left( 5\right) $. The
reciprocal space of the \emph{4D} hyperdiamond is generated by the
fundamental weight vectors $\mathbf{\omega}_{1},$ $\mathbf{\omega}_{2},$ $%
\mathbf{\omega}_{3},$ $\mathbf{\omega}_{4}$ of $SU\left( 5\right) $. Generic
wave vectors $\mathbf{k}$ in this lattice read as%
\begin{equation}
\mathbf{k}=k_{1}\mathbf{\omega}_{1}+k_{2}\mathbf{\omega}_{2}+k_{3}\mathbf{%
\omega}_{3}+k_{4}\mathbf{\omega}_{4}   \label{KW}
\end{equation}
where $k_{i}=\left( p_{i-1}-p_{i}\right) $ where $p_{i}$ is the momentum
along the $\mathbf{\lambda}_{i}$-direction and $\left( p_{i-1}-p_{i}\right) $
the momentum along the $\mathbf{\alpha}_{i}$-direction in the real 4D
hyperdiamond lattice $\mathcal{H}_{4}$. In the particular case where all the
momenta $p_{i}=\frac{4\pi N}{5d\sqrt{5}}$, we have 
\begin{equation}
\begin{tabular}{llll}
$\dsum \limits_{l=0}^{4}\mathrm{\lambda}_{l}^{\mu}e^{\pm id\frac{\sqrt{5}}{2}%
p_{l}}$ & $=$ & $e^{\pm i\frac{2\pi N}{5}}\left( \dsum \limits_{l=0}^{4}%
\mathrm{\lambda}_{l}^{\mu}\right) =0$ & .%
\end{tabular}%
\end{equation}
This property will be used later on.

\section{BBTW lattice action revisited}

\subsection{Correspondence 2D/4D}

To begin notice that a generic bond vector $\mathbf{e}_{i}$ in $\mathcal{H}%
_{4}$ links two sites in the same unit cell of the hyperdiamond as shown on
the typical coupling term $A_{\mathbf{r}_{\mathbf{n}}}B_{\mathbf{r}_{\mathbf{%
n}}+d\mathbf{e}_{i}}^{+}$. This property is quite similar to the action of
the usual $\mathrm{\gamma}^{\mu}$ matrices on \emph{4D} (Euclidean) space
time spinors. Mimicking the tight binding model of \emph{2D} graphene, BBTW
proposed in \textrm{\cite{B2}} an analogous model for \emph{4D} lattice QCD.
There construction relies on the use of the following: First, the naive
correspondence between the bond vectors $\mathbf{e}_{i}$ and the $\mathrm{%
\gamma}^{i}$ matrices%
\begin{equation}
\begin{tabular}{llllll}
$\mathbf{e}_{i}$ & $\longleftrightarrow$ & $\mathrm{\gamma}_{i}$ & , & $%
i=1,...,5$ & ,%
\end{tabular}%
\end{equation}
with%
\begin{equation}
\begin{tabular}{ll}
$-\mathbf{e}_{5}=\mathbf{e}_{1}+\mathbf{e}_{2}+\mathbf{e}_{3}+\mathbf{e}_{4}$
& , \\ 
$-\Gamma_{5}=\mathrm{\gamma}_{1}+\mathrm{\gamma}_{2}+\mathrm{\gamma}_{3}+%
\mathrm{\gamma}_{4}$ & . \\ 
& 
\end{tabular}%
\end{equation}
Recall that the four $\mathrm{\gamma}^{\mu}$ matrices satisfy the Clifford
algebra $\mathrm{\gamma}^{\mu}\mathrm{\gamma}^{\nu}+\mathrm{\gamma}^{\nu }%
\mathrm{\gamma}^{\mu}=2\delta^{\mu\nu}$, $\mathrm{\gamma}^{5}=\mathrm{\gamma 
}^{1}\mathrm{\gamma}^{2}\mathrm{\gamma}^{3}\mathrm{\gamma}^{4}$\ gives the $%
\pm$ chiralities of the two possible Weyl spinors in \emph{4D}; and $%
\Gamma_{5}$ is \textrm{precisely the matrix }$\Gamma$\textrm{\ used in the
Bori\c{c}i-Creutz fermions \cite{BC,CB}; see also eq(2.5) of ref.\cite{DS}
for a rigorous derivation using }$SU\left( 5\right) $\textrm{\ symmetry}.
Second, as in the case of 2D graphene, $\mathcal{A}_{4} $-type sites are
occupied by left $\phi_{L}=\left( \mathrm{\phi}_{\mathbf{r}}^{a}\right) $
and right $\phi_{R}=\left( \mathrm{\bar{\phi}}_{\mathbf{r}}^{\dot{a}}\right) 
$ 2-components Weyl spinors. $\mathcal{B}_{4}$-type sites are occupied by
right $\chi_{R}=\left( \mathrm{\bar{\chi}}_{\mathbf{r}+d\mathbf{e}_{i}}^{%
\dot{a}}\right) $ and left $\chi_{L}=\left( \mathrm{\chi}_{\mathbf{r}+d%
\mathbf{e}_{i}}^{a}\right) $ Weyl spinors. 
\begin{equation}
\begin{tabular}{l|l|l}
& {\small 2D graphene} & {\small 4D hyperdiamond} \\ \hline
$\left. 
\begin{array}{c}
\mathcal{A}_{4}\text{-sites at } \\ 
\mathbf{r}_{n}%
\end{array}
\right. $ & $A_{\mathbf{r}}$ & 
\begin{tabular}{ll}
$\mathrm{\phi}_{\mathbf{r}}^{a},$ & $\mathrm{\bar{\phi}}_{\mathbf{r}}^{\dot {%
a}}$%
\end{tabular}
\\ \hline
$\left. 
\begin{array}{c}
\mathcal{B}_{4}\text{-sites at } \\ 
\mathbf{r}_{n}+d\mathbf{e}_{i}%
\end{array}
\right. $ & $B_{\mathbf{r}+de_{i}}^{+}$ & 
\begin{tabular}{ll}
$\mathrm{\bar{\chi}}_{\mathbf{r}+d\mathbf{e}_{i}}^{\dot{a}},$ & $\mathrm{%
\chi }_{\mathbf{r}+d\mathbf{e}_{i}}^{a}$%
\end{tabular}
\\ \hline
couplings\QQfnmark{%
this correspondence differs from the one given by BBTW in \cite{B2}.} & $%
\left. 
\begin{array}{c}
A_{\mathbf{r}}B_{\mathbf{r}+de_{i}}^{+} \\ 
B_{\mathbf{r}+de_{i}}A_{\mathbf{r}}^{+}%
\end{array}
\right. $ & $\left. 
\begin{array}{c}
\dsum \limits_{\mu=1}^{4}\mathbf{e}_{i}^{\mu}\left( \mathrm{\phi}_{\mathbf{r}%
}^{a}\mathrm{\sigma }_{a\dot{a}}^{\mu}\text{ }\mathrm{\bar{\chi}}_{\mathbf{r}%
+d\mathbf{e}_{i}}^{\dot{a}}\right) \\ 
\dsum \limits_{\mu=1}^{4}\mathbf{e}_{i}^{\mu}\left( \mathrm{\chi}_{\mathbf{r}%
+d\mathbf{e}_{i}}^{a}\mathrm{\bar{\sigma}}_{a\dot{a}}^{\mu}\text{ }\mathrm{%
\bar{\phi}}_{\mathbf{r}}^{\dot{a}}\right)%
\end{array}
\right. $ \\ \hline
\end{tabular}%
\QQfntext{0}{
this correspondence differs from the one given by BBTW in \cite{B2}.}
\end{equation}
where the indices $a=1,2$ and $\dot{a}=\dot{1},\dot{2}$; and where summation
over $\mu$ is in the Euclidean sense. For later use, it is interesting to
notice the two following: In 2D graphene, the wave functions $A_{\mathbf{r}}$
and $B_{\mathbf{r}+de_{i}}$ describe polarized electrons in first nearest
sites of the \emph{2D} honeycomb. As the spin up and spin down components of
the electrons contribute equally, the effect of spin couplings in 2D
graphene is ignored. In the \emph{4D} hyperdiamond, we have \emph{4+4} wave
functions at each $\mathcal{A}_{4}$-type site or $\mathcal{B}_{4}$-type one.
These wave functions are given by:\newline
- the doublets $\phi^{a}=\left( \phi _{\mathbf{r}_{n}}^{1},\phi_{\mathbf{r}%
_{n}}^{2}\right) $ and $\mathrm{\bar {\phi}}_{\mathbf{r}}^{\dot{a}}=(\mathrm{%
\bar{\phi}}_{\mathbf{r}_{n}}^{\dot{1}},\mathrm{\bar{\phi}}_{\mathbf{r}_{n}}^{%
\dot{2}})$ having respectively positive and negative $\mathrm{\gamma}^{5}$
chirality, these are the $(\frac{1}{2},0)$\ and $(0,\frac{1}{2})$
representations of the $SO\left( 4\right) \simeq SU\left( 2\right) \times
SU\left( 2\right) .$ \newline
- the doublets $\bar{\chi}_{\mathbf{r}+d\mathbf{e}_{i}}^{\dot{a}}=(\bar{\chi 
}_{\mathbf{r}+d\mathbf{e}_{i}}^{\dot{1}},\bar{\chi}_{\mathbf{r}+d\mathbf{e}%
_{i}}^{\dot{2}})$ and $\mathrm{\chi}_{\mathbf{r}+d\mathbf{e}_{i}}^{a}=(%
\mathrm{\chi}_{\mathbf{r}+d\mathbf{e}_{i}}^{1},\mathrm{\chi}_{\mathbf{r}+d%
\mathbf{e}_{i}}^{2})$ having respectively negative and positive $\mathrm{%
\gamma}^{5}$ chirality.\newline
By mimicking the 2D graphene study, we expect therefore to have \emph{4}
kinds of polarized particles together with the \emph{4} corresponding
"holes" as shown on the typical tight binding couplings%
\begin{equation}
\begin{tabular}{llll}
$\mathbf{e}_{i}^{\mu}\mathrm{\sigma}_{1\dot{1}}^{\mu}\left( \mathrm{\phi }_{%
\mathbf{r}}^{1}\text{ }\mathrm{\bar{\chi}}_{\mathbf{r}+d\mathbf{e}_{i}}^{%
\dot{1}}\right) $ & , & $\mathbf{e}_{i}^{\mu}\mathrm{\sigma}_{2\dot{2}%
}^{\mu}\left( \mathrm{\phi}_{\mathbf{r}}^{2}\text{ }\mathrm{\bar{\chi}}_{%
\mathbf{r}+d\mathbf{e}_{i}}^{\dot{2}}\right) $ &  \\ 
&  &  &  \\ 
$\mathbf{e}_{i}^{\mu}\mathrm{\bar{\sigma}}_{1\dot{1}}^{\mu}\left( \mathrm{%
\chi}_{\mathbf{r}+d\mathbf{e}_{i}}^{1}\text{ }\mathrm{\bar{\phi}}_{\mathbf{r}%
}^{\dot{1}}\right) $ & , & $\mathbf{e}_{i}^{\mu}\mathrm{\bar {\sigma}}_{2%
\dot{2}}^{\mu}\left( \mathrm{\chi}_{\mathbf{r}+d\mathbf{e}_{i}}^{2}\text{ }%
\mathrm{\bar{\phi}}_{\mathbf{r}}^{\dot{2}}\right) $ & 
\end{tabular}%
\end{equation}

\subsection{Building the action}

Following \textrm{\cite{B2}}, the BBTW action is a naive lattice QCD action
preserving the symmetries of $\mathcal{H}_{4}$. To describe the spinor
structures of the lattice fermions, one considers \emph{4D} space time Dirac
spinors together with the following $\mathrm{\gamma}^{\mu}$ matrices
realizations,%
\begin{equation}
\begin{tabular}{llllll}
$\mathrm{\gamma}^{1}=\mathrm{\tau}^{1}\otimes\mathrm{\sigma}^{1}$ & , & $%
\mathrm{\gamma}^{2}=\mathrm{\tau}^{1}\otimes\mathrm{\sigma}^{2}$ & , & $%
\mathrm{\gamma}^{3}=\mathrm{\tau}^{1}\otimes\mathrm{\sigma}^{3}$ & , \\ 
$\mathrm{\gamma}^{4}=\mathrm{\tau}^{2}\otimes\mathrm{I}_{2}$ & , & $\mathrm{%
\gamma}^{5}=\mathrm{\tau}^{3}\otimes\mathrm{I}_{2}$ & , &  & 
\end{tabular}%
\end{equation}
where the $\tau^{i}$'s are the Pauli matrices acting on the sublattice
structure of the hyperdiamond lattice $\mathcal{H}_{4}$, 
\begin{equation}
\begin{tabular}{llll}
$\mathrm{\tau}^{1}=\left( 
\begin{array}{cc}
0 & 1 \\ 
1 & 0%
\end{array}
\right) ,$ & $\mathrm{\tau}^{2}=\left( 
\begin{array}{cc}
0 & -i \\ 
i & 0%
\end{array}
\right) ,$ & $\mathrm{\tau}^{3}=\left( 
\begin{array}{cc}
1 & 0 \\ 
0 & -1%
\end{array}
\right) .$ & 
\end{tabular}%
\end{equation}
The $2\times2$ matrices $\mathrm{\sigma}^{i}$ satisfy as well the Clifford
algebra $\mathrm{\sigma}^{i}\mathrm{\sigma}^{j}+\mathrm{\sigma}^{j}\mathrm{%
\sigma}^{i}=2\delta^{ij}\mathrm{I}_{2}$ and act through the coupling of left 
$\phi_{L}$ (resp. $\phi_{R}$) and right $\chi_{R}$ (resp. left $\chi_{L}$)
2-components Weyl spinors at neighboring $\mathcal{A}_{4}$- and $\mathcal{B}%
_{4}$- sites%
\begin{equation}
\mathrm{\phi}_{\mathbf{r}}^{a}\mathrm{\sigma}_{a\dot{a}}^{\mu}\text{ }%
\mathrm{\bar{\chi}}_{\mathbf{r}+d\frac{\sqrt{5}}{2}\mathbf{\lambda}_{i}}^{%
\dot{a}}-\mathrm{\chi}_{\mathbf{r}}^{a}\mathrm{\bar{\sigma}}_{a\dot{a}}^{\mu}%
\text{ }\mathrm{\bar{\phi}}_{\mathbf{r-}d\frac{\sqrt{5}}{2}\mathbf{\lambda}%
_{i}}^{\dot{a}}=\left( \mathrm{\phi}_{\mathbf{r}}\mathrm{\sigma}^{\mu}%
\mathrm{\bar{\chi}}_{\mathbf{r}+d\frac{\sqrt{5}}{2}\mathbf{\lambda}_{i}}-%
\mathrm{\chi}_{\mathbf{r}}\mathrm{\bar{\sigma}}^{\mu }\mathrm{\bar{\phi}}_{%
\mathbf{r}-d\frac{\sqrt{5}}{2}\mathbf{\lambda}_{i}}\right)
\end{equation}
where $\mathrm{\sigma}^{\mu}=\left( \mathrm{\sigma}^{1},\mathrm{\sigma}^{2},%
\mathrm{\sigma}^{3},+i\mathrm{I}_{2}\right) $ and $\mathrm{\bar{\sigma}}%
^{\mu}=\left( \mathrm{\sigma}^{1},\mathrm{\sigma}^{2},\mathrm{\sigma}^{3},-i%
\mathrm{I}_{2}\right) $. For later use, it is interesting to set%
\begin{equation}
\begin{tabular}{lll}
$\mathrm{\sigma}^{\mu}\mathrm{.}\mathbf{e}_{1}^{\mu}$ & $=\frac{\sqrt{5}}{4}%
\mathrm{\sigma}^{1}+\frac{\sqrt{5}}{4}\mathrm{\sigma}^{2}+\frac{\sqrt{5}}{4}%
\mathrm{\sigma}^{3}+\frac{i}{4}\mathrm{I}_{2}$ & , \\ 
$\mathrm{\bar{\sigma}}^{\mu}\mathrm{.}\mathbf{e}_{1}^{\mu}$ & $=\frac{\sqrt {%
5}}{4}\mathrm{\sigma}^{1}+\frac{\sqrt{5}}{4}\mathrm{\sigma}^{2}+\frac {\sqrt{%
5}}{4}\mathrm{\sigma}^{3}-\frac{i}{4}\mathrm{I}_{2}$ & ,%
\end{tabular}%
\end{equation}
and similar relations for the other $\mathrm{\sigma.}\mathbf{e}_{i}$ and $%
\mathrm{\bar{\sigma}.}\mathbf{e}_{i}$.\ \newline
Now extending the tight binding model of \emph{2D }graphene to the \emph{4D}
hyperdiamond; and using the weight vectors $\mathbf{\lambda}_{i}$ instead of 
$\mathbf{e}_{i}$, we can build a free fermion action on the lattice $%
\mathcal{H}_{4}$ by attaching a two-component left-handed spinor $\mathrm{%
\phi}^{a}\left( \mathbf{r}\right) $ and right-handed spinor $\mathrm{\bar{%
\phi}}_{\mathbf{r}}^{\dot{a}}$ to each $\mathcal{A}_{4}$-node $\mathbf{r}$,
and a right-handed spinor $\mathrm{\bar {\chi}}_{\mathbf{r}+d\frac{\sqrt{5}}{%
2}\mathbf{\lambda}_{i}}^{\dot{a}}$ and left-handed spinor $\mathrm{\chi}_{%
\mathbf{r}+d\frac{\sqrt{5}}{2}\mathbf{\lambda}_{i}}^{a}$ to every $\mathcal{B%
}_{4}$-node at $\mathbf{r}+d\frac{\sqrt{5}}{2}\mathbf{\lambda}_{i}$. The
action, describing hopping to first nearest-neighbor sites with equal
probabilities in all five directions $\mathbf{\lambda}_{i}$, reads as
follows:.%
\begin{equation}
\begin{tabular}{llll}
$\mathcal{S}_{_{{\small BBTW}}}$ & $=$ & $\dsum \limits_{\mathbf{r}}\dsum
\limits_{i=0}^{4}\left( \mathrm{\phi}_{\mathbf{r}}\mathrm{\sigma}^{\mu}%
\mathrm{\bar{\chi}}_{\mathbf{r+}d\frac{\sqrt{5}}{2}\mathbf{\lambda}_{i}}-%
\mathrm{\chi }_{\mathbf{r}}\mathrm{\bar{\sigma}}^{\mu}\mathrm{\bar{\phi}}_{%
\mathbf{r-}d\frac{\sqrt{5}}{2}\mathbf{\lambda}_{i}}\right) \mathrm{\lambda}%
_{i}^{\mu}$ & .%
\end{tabular}
\label{AC}
\end{equation}
Clearly this action is invariant under the following discrete
transformations 
\begin{equation}
\begin{tabular}{lllllll}
$\mathrm{\sigma}^{\mu}\mathrm{\bar{\xi}}_{\mathbf{r}\pm d\frac{\sqrt{5}}{2}%
\mathbf{\lambda}_{i}}$ & $\longrightarrow$ & $\mathrm{\sigma}^{\nu }\mathrm{%
\bar{\xi}}_{\mathbf{r}\pm d\frac{\sqrt{5}}{2}\mathbf{\lambda}_{j}}\left( 
\mathcal{O}_{ji}^{T}\right) _{\nu}^{\mu}$ & , & $\mathrm{\lambda }_{i}^{\mu}$
& $\longrightarrow$ & $\left( \mathcal{O}_{ji}\right) _{\rho }^{\mu}\mathrm{%
\lambda}_{j}^{\rho}.$%
\end{tabular}%
\end{equation}
Expanding the various spinorial fields $\mathrm{\xi}_{\mathbf{r\pm v}}$ in
Fourier sums as $\int\frac{d^{4}k}{\left( 2\pi\right) ^{4}}e^{-i\mathbf{k.r}%
}\left( e^{\mp i\mathbf{k.v}}\mathrm{\xi}_{\mathbf{k}}\right) $ with $%
\mathbf{k}$ standing for a generic wave vector in $\mathcal{H}_{4}^{\ast}$,
we can put the field action $\mathcal{S}_{_{{\small BBTW}}}$ into the form 
\begin{equation}
\begin{tabular}{llll}
$\mathcal{S}_{_{{\small BBTW}}}$ & $=$ & $i\dsum \limits_{\mathbf{k}}\left( 
\mathrm{\bar{\phi}}_{\mathbf{k}},\mathrm{\bar{\chi}}_{\mathbf{k}}\right)
\left( 
\begin{array}{cc}
0 & -iD \\ 
i\bar{D} & 0%
\end{array}
\right) \left( 
\begin{array}{c}
\mathrm{\phi}_{\mathbf{k}} \\ 
\mathrm{\chi}_{\mathbf{k}}%
\end{array}
\right) $ & 
\end{tabular}
\label{D1}
\end{equation}
where we have set%
\begin{equation}
\begin{tabular}{llll}
$D$ & $=\dsum \limits_{l=0}^{4}D_{l}e^{id\frac{\sqrt{5}}{2}\mathbf{k}.%
\mathrm{\lambda}_{l}}$ & $=\dsum \limits_{\mu=1}^{4}\mathrm{\sigma}%
^{\mu}\left( \dsum \limits_{l=0}^{4}\mathrm{\lambda}_{l}^{\mu}e^{id\frac{%
\sqrt{5}}{2}\mathbf{k}.\mathrm{\lambda }_{l}}\right) $ & ,%
\end{tabular}
\label{D2}
\end{equation}
with%
\begin{equation}
\begin{tabular}{ll}
$D_{l}=\dsum \limits_{\mu=1}^{4}\mathrm{\sigma}^{\mu}\mathrm{\lambda}%
_{l}^{\mu}=\left( 
\begin{array}{cc}
\mathrm{\lambda}_{l}^{3}+i\mathrm{\lambda}_{l}^{4} & \mathrm{\lambda}%
_{l}^{1}-i\mathrm{\lambda}_{l}^{2} \\ 
\mathrm{\lambda}_{l}^{1}+i\mathrm{\lambda}_{l}^{2} & \mathrm{\lambda}%
_{l}^{3}-i\mathrm{\lambda}_{l}^{4}%
\end{array}
\right) $ & ,%
\end{tabular}%
\end{equation}
and $p_{l}=\mathbf{k}.\mathrm{\lambda}_{l}=\sum_{\mu}\mathbf{k}_{\mu }%
\mathrm{\lambda}_{l}^{\mu}$. Similarly we have%
\begin{equation}
\begin{tabular}{llll}
$\bar{D}$ & $=\dsum \limits_{l=0}^{4}\bar{D}_{l}e^{-id\frac{\sqrt{5}}{2}%
\mathbf{k}.\mathrm{\lambda}_{l}}$ & $=\dsum \limits_{\mu=1}^{4}\mathrm{\bar{%
\sigma}}^{\mu}\left( \dsum \limits_{l=0}^{4}\mathrm{\lambda}_{l}^{\mu}e^{-id%
\frac{\sqrt{5}}{2}\mathbf{k}.\mathrm{\lambda }_{l}}\right) $ & ,%
\end{tabular}
\label{2D}
\end{equation}
We end this subsection by making \emph{3} remarks; the first one deals with
the continuous limit; the second one regards the zeros of the Dirac operator
and the third concerns the link with the Creutz fermions. In the continuous
limit where the lattice parameter $d\rightarrow0$, we have%
\begin{equation}
\begin{tabular}{llll}
$\dsum \limits_{l=0}^{4}\mathrm{\lambda}_{l}^{\mu}e^{\pm id\frac{\sqrt{5}}{2}%
\mathbf{k}.\mathrm{\lambda}_{l}}$ & $\rightarrow$ & $\left( \dsum
\limits_{l=0}^{4}\mathrm{\lambda}_{l}^{\mu}\right) \pm i\frac{d\sqrt{5}}{2}%
\left[ \dsum \limits_{l=0}^{4}\mathrm{\lambda}_{l}^{\mu}\left( \mathbf{k}.%
\mathrm{\lambda}_{l}\right) \right] +...$ & .%
\end{tabular}%
\end{equation}
Moreover, since $\sum_{l=0}^{4}\mathrm{\lambda}_{l}^{\mu}=0$ and because of
the identity $\sum_{l=0}^{4}\mathrm{\lambda}_{l}^{\mu}\left( \mathbf{k}.%
\mathrm{\lambda}_{l}\right) =\mathbf{k}^{\mu}$ following from eqs(\ref{X}-%
\ref{Y}), this limit reduces to%
\begin{equation}
\begin{tabular}{llll}
$\dsum \limits_{l=0}^{4}\mathrm{\lambda}_{l}^{\mu}e^{\pm id\frac{\sqrt{5}}{2}%
\mathbf{k}.\mathrm{\lambda}_{l}}$ & $\rightarrow$ & $\pm i\frac{d\sqrt{5}}{2}%
\mathbf{k}^{\mu}+...$ & .%
\end{tabular}%
\end{equation}
So we have%
\begin{equation}
\begin{tabular}{llll}
$D\longrightarrow i\frac{d\sqrt{5}}{2}\dsum \limits_{\mu=1}^{4}\mathrm{\sigma%
}^{\mu}\mathbf{k}_{\mu}$ & , & $\bar{D}\longrightarrow -i\frac{d\sqrt{5}}{2}%
\dsum \limits_{\mu=1}^{4}\mathrm{\bar{\sigma}}^{\mu}\mathbf{k}_{\mu}$ & .%
\end{tabular}%
\end{equation}
The operators $D$ and $\bar{D}$ have zeros for wave vectors $\mathbf{k}$
satisfying the following constraint relation 
\begin{equation}
\begin{tabular}{llll}
$\mathbf{k}.\mathrm{\lambda}_{l}$ & $=$ & $\frac{4\pi N}{5d\sqrt{5}}$ & ,%
\end{tabular}
\label{NL}
\end{equation}
with $N$\ an arbitrary integer. The point is that for these values, the
phases $e^{id\frac{\sqrt{5}}{2}\mathbf{k}.\mathrm{\lambda}_{l}}=e^{i\mathbf{%
\varphi}}$ and the operators $D$ and $\bar{D}$ get reduced to 
\begin{equation}
\begin{tabular}{llllll}
$D=$ & $e^{i\mathbf{\varphi}}\dsum \limits_{\mu=1}^{4}\mathrm{\sigma}%
^{\mu}\left( \dsum \limits_{l=0}^{4}\mathrm{\lambda}_{l}^{\mu}\right) $ & ,
& $\bar{D}=$ & $e^{-i\mathbf{\varphi }}\dsum \limits_{\mu=1}^{4}\mathrm{\bar{%
\sigma}}^{\mu}\left( \dsum \limits_{l=0}^{4}\mathrm{\lambda}%
_{l}^{\mu}\right) $ & 
\end{tabular}%
\end{equation}
which vanish identically due to the property $\sum_{l=0}^{4}\mathrm{\lambda }%
_{l}^{\mu}=0$. Following \textrm{\cite{B4,B3}}, the Dirac operator (\ref{D1}%
) in the Creutz lattice model reads as follows,%
\begin{equation}
\left( 
\begin{array}{cc}
0 & z \\ 
z^{\ast} & 0%
\end{array}
\right)
\end{equation}
where $z=\theta_{0}I+i\theta_{1}\mathrm{\sigma}^{1}+i\theta_{2}\mathrm{%
\sigma }^{2}+i\theta_{3}\mathrm{\sigma}^{3}$ with%
\begin{equation}
\begin{tabular}{ll}
$\theta_{1}=$ & $\sin p_{1}+\sin p_{2}-\sin p_{3}-\sin p_{4}$ \\ 
$\theta_{2}=$ & $\sin p_{1}-\sin p_{2}-\sin p_{3}+\sin p_{4}$ \\ 
$\theta_{3}=$ & $\sin p_{1}-\sin p_{2}+\sin p_{3}-\sin p_{4}$ \\ 
$\theta_{0}=$ & $B\left( 4C-\cos p_{1}-\cos p_{2}-\cos p_{3}-\cos
p_{4}\right) $%
\end{tabular}
\label{te}
\end{equation}
and B and C two real parameters. In the Creutz lattice model, the zero
energy states correspond to $z=0$; this leads to the constraints $%
\theta_{i}=0$ which are solved by taking one of the momenta as $p_{1}=p$ and
the others as $p_{i}=p$ or $\pi-p$. To make contact with our construction,
the analogous of eqs(\ref{te}) are given by:%
\begin{equation}
\begin{tabular}{ll}
&  \\ 
$\theta_{1}=$ & $\mathrm{\lambda}_{0}^{1}e^{-id\frac{\sqrt{5}}{2}p_{0}}+%
\mathrm{\lambda}_{1}^{1}e^{-id\frac{\sqrt{5}}{2}p_{1}}+\mathrm{\lambda}%
_{2}^{1}e^{-id\frac{\sqrt{5}}{2}p_{2}}+\mathrm{\lambda}_{3}^{1}e^{-id\frac {%
\sqrt{5}}{2}p_{3}}+\mathrm{\lambda}_{4}^{1}e^{-id\frac{\sqrt{5}}{2}p_{4}}$
\\ 
$\theta_{2}=$ & $\mathrm{\lambda}_{0}^{2}e^{-id\frac{\sqrt{5}}{2}p_{0}}+%
\mathrm{\lambda}_{1}^{2}e^{-id\frac{\sqrt{5}}{2}p_{1}}+\mathrm{\lambda}%
_{2}^{2}e^{-id\frac{\sqrt{5}}{2}p_{2}}+\mathrm{\lambda}_{3}^{2}e^{-id\frac {%
\sqrt{5}}{2}p_{3}}+\mathrm{\lambda}_{4}^{2}e^{-id\frac{\sqrt{5}}{2}p_{4}}$
\\ 
$\theta_{3}=$ & $\mathrm{\lambda}_{0}^{3}e^{-id\frac{\sqrt{5}}{2}p_{0}}+%
\mathrm{\lambda}_{1}^{3}e^{-id\frac{\sqrt{5}}{2}p_{1}}+\mathrm{\lambda}%
_{2}^{3}e^{-id\frac{\sqrt{5}}{2}p_{2}}+\mathrm{\lambda}_{3}^{3}e^{-id\frac {%
\sqrt{5}}{2}p_{3}}+\mathrm{\lambda}_{4}^{3}e^{-id\frac{\sqrt{5}}{2}p_{4}}$
\\ 
$\theta_{0}=$ & $\mathrm{\lambda}_{0}^{4}e^{-id\frac{\sqrt{5}}{2}p_{0}}+%
\mathrm{\lambda}_{1}^{4}e^{-id\frac{\sqrt{5}}{2}p_{1}}+\mathrm{\lambda}%
_{2}^{4}e^{-id\frac{\sqrt{5}}{2}p_{2}}+\mathrm{\lambda}_{3}^{4}e^{-id\frac {%
\sqrt{5}}{2}p_{3}}+\mathrm{\lambda}_{4}^{4}e^{-id\frac{\sqrt{5}}{2}p_{4}}$
\\ 
& 
\end{tabular}
\label{et}
\end{equation}
where $p_{l}=\mathbf{k}.\mathrm{\lambda}_{l}$. These relations are complex
and are, in some sense, more general than the Creutz ones (\ref{te}). The
zeros of these solutions requires $e^{id\frac{\sqrt{5}}{2}p_{i}}=e^{i\varphi}
$ $\forall$ $l=0,1,2,3,4$ as anticipated in (\ref{cst}).

\section{Energy dispersion and zero modes}

To get the dispersion energy relations of the \emph{4} waves components $%
\mathrm{\phi}_{\mathbf{k}}^{1}$, $\mathrm{\phi}_{\mathbf{k}}^{2}$, $\mathrm{%
\chi}_{\mathbf{k}}^{1},$ $\mathrm{\chi}_{\mathbf{k}}^{2}$ and their
corresponding \emph{4} holes, one has to solve the eigenvalues of the Dirac
operator (\ref{D1}). To that purpose, we first write the \emph{4}%
-dimensional wave equation as follows,%
\begin{equation}
\begin{tabular}{llll}
$\left( 
\begin{array}{cc}
0 & -iD \\ 
i\bar{D} & 0%
\end{array}
\right) \left( 
\begin{array}{c}
\mathrm{\phi}_{\mathbf{k}} \\ 
\mathrm{\chi}_{\mathbf{k}}%
\end{array}
\right) $ & $=$ & $E\left( 
\begin{array}{c}
\mathrm{\phi}_{\mathbf{k}} \\ 
\mathrm{\chi}_{\mathbf{k}}%
\end{array}
\right) $ & ,%
\end{tabular}
\label{DE}
\end{equation}
where $\mathrm{\phi}_{\mathbf{k}}=\left( \mathrm{\phi}_{\mathbf{k}}^{1},%
\mathrm{\phi}_{\mathbf{k}}^{2}\right) $, $\mathrm{\chi}_{\mathbf{k}}=\left( 
\mathrm{\chi}_{\mathbf{k}}^{1},\mathrm{\chi}_{\mathbf{k}}^{2}\right) $ are
Weyl spinors and where the $2\times2$ matrices $D$, $\bar{D}$ are as in eqs(%
\ref{D2},\ref{2D}). Then determine the eigenstates and eigenvalues of the $%
2\times2$ Dirac operator matrix by solving the following characteristic
equation,%
\begin{equation}
\begin{tabular}{lll}
$\det\left( 
\begin{array}{cccc}
-E & 0 & D_{11} & D_{12} \\ 
0 & -E & D_{21} & D_{22} \\ 
\bar{D}_{11} & \bar{D}_{21} & -E & 0 \\ 
\bar{D}_{12} & \bar{D}_{22} & 0 & -E%
\end{array}
\right) $ & $=0$ & 
\end{tabular}%
\end{equation}
from which one can learn the four dispersion energy eigenvalues $E_{1}\left( 
\mathbf{k}\right) $, $E_{2}\left( \mathbf{k}\right) $, $E_{3}\left( \mathbf{k%
}\right) $, $E_{4}\left( \mathbf{k}\right) $ and therefore their zeros.

\subsection{Computing the energy dispersion}

An interesting way to do these calculations is to act on (\ref{DE}) once
more by the Dirac operator to bring it to the following diagonal form%
\begin{equation}
\begin{tabular}{llll}
$\left( 
\begin{array}{cc}
D\bar{D} & 0 \\ 
0 & D\bar{D}%
\end{array}
\right) \left( 
\begin{array}{c}
\mathrm{\phi}_{\mathbf{k}} \\ 
\mathrm{\chi}_{\mathbf{k}}%
\end{array}
\right) $ & $=$ & $E^{2}\left( 
\begin{array}{c}
\mathrm{\phi}_{\mathbf{k}} \\ 
\mathrm{\chi}_{\mathbf{k}}%
\end{array}
\right) $ & .%
\end{tabular}%
\end{equation}
Then solve separately the eigenvalues problem of the 2-dimensional equations 
$D\bar{D}\mathrm{\phi}_{\mathbf{k}}=E^{2}\mathrm{\phi}_{\mathbf{k}}$ and $%
\bar{D}D\mathrm{\chi}_{\mathbf{k}}=E^{2}\mathrm{\chi}_{\mathbf{k}}$. To do
so, it is useful to set 
\begin{equation}
\begin{tabular}{llll}
$u\left( \mathbf{k}\right) =\mathrm{\vartheta}^{1}+i\mathrm{\vartheta}^{2}$
& , & $v\left( \mathbf{k}\right) =\mathrm{\vartheta}^{3}+i\mathrm{\vartheta }%
^{4}$ & 
\end{tabular}
\label{uv}
\end{equation}
with 
\begin{equation}
\begin{tabular}{lll}
$\mathrm{\vartheta}^{\mu}=\dsum \limits_{l=0}^{4}\mathrm{\lambda}%
_{l}^{\mu}e^{id\frac{\sqrt{5}}{2}\mathbf{k}.\mathrm{\lambda }_{l}}$ & , & $%
\mu=1,2,3,4.$%
\end{tabular}%
\end{equation}
Notice that in the continuous limit, we have%
\begin{equation}
\begin{tabular}{llll}
$\mathrm{\vartheta}^{\mu}$ & $\longrightarrow$ & $id\frac{\sqrt{5}}{2}%
\mathbf{k}^{\mu}$ & , \\ 
$u\left( \mathbf{k}\right) $ & $\longrightarrow$ & $id\frac{\sqrt{5}}{2}%
\left( \mathbf{k}^{1}+i\mathbf{k}^{2}\right) $ & , \\ 
$v\left( \mathbf{k}\right) $ & $\longrightarrow$ & $id\frac{\sqrt{5}}{2}%
\left( \mathbf{k}^{3}+i\mathbf{k}^{4}\right) $ & .%
\end{tabular}%
\end{equation}
Substituting (\ref{uv}) back into (\ref{D2}) and (\ref{2D}), we obtain the
following expressions,%
\begin{equation}
\begin{tabular}{lll}
$D\bar{D}=\left( 
\begin{array}{cc}
\left\vert u\right\vert ^{2}+\left\vert v\right\vert ^{2} & 2\bar{u}v \\ 
2u\bar{v} & \left\vert u\right\vert ^{2}+\left\vert v\right\vert ^{2}%
\end{array}
\right) ,$ &  & 
\end{tabular}%
\end{equation}
and%
\begin{equation}
\begin{tabular}{lll}
$\bar{D}D=\left( 
\begin{array}{cc}
\left\vert u\right\vert ^{2}+\left\vert v\right\vert ^{2} & 2\bar{u}\bar{v}
\\ 
2uv & \left\vert u\right\vert ^{2}+\left\vert v\right\vert ^{2}%
\end{array}
\right) .$ &  & 
\end{tabular}%
\end{equation}
By solving the characteristic equations of these $2\times2$ matrix
operators, we get the following eigenstates $\mathrm{\phi}_{\mathbf{k}%
}^{a\prime}$, $\mathrm{\chi}_{\mathbf{k}}^{a\prime}$ with their
corresponding eigenvalues $E_{\pm}^{2}$,

\begin{equation}
\begin{tabular}{lll|ll}
\multicolumn{2}{l}{\small \ \ \ \ \ \ \ \ \ eigenstates} &  &  & {\small \ \
\ \ \ \ \ \ \ eigenvalues} \\ \hline
&  &  &  &  \\ 
$\mathrm{\phi}_{\mathbf{k}}^{1\prime}=$ & $\sqrt{\frac{v\bar{u}}{2\left\vert
u\right\vert \left\vert v\right\vert }}\mathrm{\phi}_{\mathbf{k}}^{1}+\sqrt{%
\frac{u\bar{v}}{2\left\vert u\right\vert \left\vert v\right\vert }}\mathrm{%
\phi}_{\mathbf{k}}^{2}$ &  &  & $E_{+}^{2}=\left\vert u\right\vert
^{2}+\left\vert v\right\vert ^{2}+2\left\vert u\right\vert \left\vert
v\right\vert $ \\ 
&  &  &  &  \\ 
$\mathrm{\phi}_{\mathbf{k}}^{2\prime}=$ & $-\sqrt{\frac{v\bar{u}}{%
2\left\vert u\right\vert \left\vert v\right\vert }}\mathrm{\phi}_{\mathbf{k}%
}^{1}+\sqrt{\frac{u\bar{v}}{2\left\vert u\right\vert \left\vert v\right\vert 
}}\mathrm{\phi}_{\mathbf{k}}^{2}$ &  &  & $E_{-}^{2}=\left\vert u\right\vert
^{2}+\left\vert v\right\vert ^{2}-2\left\vert u\right\vert \left\vert
v\right\vert $ \\ 
&  &  &  &  \\ \hline
\end{tabular}%
\end{equation}
and

\begin{equation}
\begin{tabular}{lll|ll}
\multicolumn{2}{l}{\small \ \ \ \ \ \ \ \ \ eigenstates} &  &  & {\small \ \
\ \ \ \ \ \ \ eigenvalues} \\ \hline
&  &  &  &  \\ 
$\mathrm{\chi}_{\mathbf{k}}^{1\prime}=$ & $\sqrt{\frac{\bar{u}\bar{v}}{%
2\left\vert u\right\vert \left\vert v\right\vert }}\mathrm{\chi}_{\mathbf{k}%
}^{1}+\sqrt{\frac{uv}{2\left\vert u\right\vert \left\vert v\right\vert }}%
\mathrm{\chi}_{\mathbf{k}}^{2}$ &  &  & $E_{+}^{2}=\left\vert u\right\vert
^{2}+\left\vert v\right\vert ^{2}+2\left\vert u\right\vert \left\vert
v\right\vert $ \\ 
&  &  &  &  \\ 
$\mathrm{\chi}_{\mathbf{k}}^{2\prime}=$ & $-\sqrt{\frac{\bar{u}\bar{v}}{%
2\left\vert u\right\vert \left\vert v\right\vert }}\mathrm{\chi}_{\mathbf{k}%
}^{1}+\sqrt{\frac{uv}{2\left\vert u\right\vert \left\vert v\right\vert }}%
\mathrm{\chi}_{\mathbf{k}}^{2}$ &  &  & $E_{-}^{2}=\left\vert u\right\vert
^{2}+\left\vert v\right\vert ^{2}-2\left\vert u\right\vert \left\vert
v\right\vert $ \\ 
&  &  &  &  \\ \hline
\end{tabular}%
\end{equation}

\ \ \ \newline
By taking square roots of $E_{\pm}^{2}$, we obtain \emph{2} positive and 
\emph{2} negative dispersion energies; these are 
\begin{equation}
\begin{tabular}{ll}
$E_{\pm}=+\sqrt{\left( \left\vert u\right\vert \pm\left\vert v\right\vert
\right) ^{2}}$ & 
\end{tabular}%
\end{equation}
which correspond to particles; and 
\begin{equation}
\begin{tabular}{ll}
$E_{\pm}^{\ast}=-\sqrt{\left( \left\vert u\right\vert \pm\left\vert
v\right\vert \right) ^{2}}$ & 
\end{tabular}%
\end{equation}
corresponding to the associated holes.

\subsection{Determining the zeros of $E_{\pm}$ and $E_{\pm}^{\ast}$}

From the above energy dispersion relations, one sees that the zero modes are
of two kinds as listed here below\textrm{:}

\emph{zeros of both }$E_{+}^{2}=0,$\emph{\ }$E_{-}^{2}=0$\newline
They are given by those wave vectors $\mathbf{K}_{F}$ solving the constraint
relations $u\left( \mathbf{K}_{F}\right) =v\left( \mathbf{K}_{F}\right) =0$
which can be also put in the form%
\begin{equation}
\begin{tabular}{lll}
$\mathrm{\lambda}_{0}^{\mu}e^{id\frac{\sqrt{5}}{2}\mathbf{K}_{F}.\mathrm{%
\lambda}_{0}}+\mathrm{\lambda}_{1}^{\mu}e^{id\frac{\sqrt{5}}{2}\mathbf{K}%
_{F}.\mathrm{\lambda}_{1}}+$ &  &  \\ 
$+\mathrm{\lambda}_{2}^{\mu}e^{id\frac{\sqrt{5}}{2}\mathbf{K}_{F}.\mathrm{%
\lambda}_{2}}+\mathrm{\lambda}_{3}^{\mu}e^{id\frac{\sqrt{5}}{2}\mathbf{K}%
_{F}.\mathrm{\lambda}_{3}}+\mathrm{\lambda}_{4}^{\mu}e^{id\frac{\sqrt{5}}{2}%
\mathbf{K}_{F}.\mathrm{\lambda}_{4}}$ & $=0$ & 
\end{tabular}%
\end{equation}
for all values of $\mu=1,2,3,4$; or equivalently like 
\begin{equation}
\begin{array}{cc}
d\frac{\sqrt{5}}{2}\mathbf{K}_{F}.\mathrm{\lambda}_{l} & =\frac{2\pi}{5}%
N+2\pi N_{l}.%
\end{array}%
\end{equation}
The solutions of these constraint equations have been studied in section 3;
they are precisely given by eqs(\ref{S2}-\ref{Y}). Now, setting $\mathbf{k}=%
\mathbf{K}_{F}+\mathbf{q}$ with small $q=\left\Vert \mathbf{q}\right\Vert $
and expanding $D$ and $\bar{D}$, eq(\ref{DE}) gets reduced to 
\begin{equation}
\begin{tabular}{llll}
$\frac{d\sqrt{5}}{2}\dsum \limits_{\mu=1}^{4}\mathbf{q}_{\mu}\left( 
\begin{array}{cc}
0 & \mathrm{\sigma}^{\mu} \\ 
\mathrm{\bar{\sigma}}^{\mu} & 0%
\end{array}
\right) \left( 
\begin{array}{c}
\mathrm{\phi}_{\mathbf{k}} \\ 
\mathrm{\chi}_{\mathbf{k}}%
\end{array}
\right) $ & $=$ & $E\left( 
\begin{array}{c}
\mathrm{\phi}_{\mathbf{k}} \\ 
\mathrm{\chi}_{\mathbf{k}}%
\end{array}
\right) $ & .%
\end{tabular}%
\end{equation}

\emph{case }$E_{-}^{2}=0$\emph{\ but }$E_{+}^{2}=E_{+\min}^{2}\neq0$ \newline
These minima are given by those wave vectors $\mathbf{K=k}_{\min}$ solving
the following constraint relation $\left\vert u\left( \mathbf{K}\right)
\right\vert =\left\vert v\left( \mathbf{K}\right) \right\vert $ or
equivalently%
\begin{equation}
\begin{tabular}{lll}
$\dsum \limits_{m,n=0}^{4}\left( \mathrm{\lambda}_{m}^{1}+i\mathrm{\lambda}%
_{m}^{2}\right) \left( \mathrm{\lambda}_{n}^{1}-i\mathrm{\lambda}%
_{n}^{2}\right) e^{id\frac{\sqrt {5}}{2}\mathbf{K}.\mathrm{\beta}_{mn}}=$ & 
&  \\ 
$\dsum \limits_{m,n=0}^{4}\left( \mathrm{\lambda}_{m}^{3}+i\mathrm{\lambda}%
_{m}^{4}\right) \left( \mathrm{\lambda}_{n}^{3}-i\mathrm{\lambda}%
_{n}^{4}\right) e^{id\frac{\sqrt {5}}{2}\mathbf{K}.\mathrm{\beta}_{mn}}$ & .
& 
\end{tabular}%
\end{equation}
Expanding this equality, we get the following condition on the wave vector,%
\begin{equation}
\begin{tabular}{ll}
$\dsum \limits_{m,n=0}^{4}\mathcal{A}_{nm}\cos\left( d\frac{\sqrt{5}}{2}%
\mathbf{K}.\mathrm{\beta}_{mn}\right) \left[ \tan\left( d\frac{\sqrt{5}}{2}%
\mathbf{K}.\mathrm{\beta }_{mn}\right) -\frac{\mathcal{B}_{nm}}{\mathcal{A}%
_{nm}}\right] =0$ & ,%
\end{tabular}%
\end{equation}
with%
\begin{equation}
\begin{tabular}{llll}
$\mathcal{A}_{nm}$ & $=$ & $\left( \mathrm{\lambda}_{n}^{1}\mathrm{\lambda }%
_{m}^{2}-\mathrm{\lambda}_{m}^{1}\mathrm{\lambda}_{n}^{2}\right) -\left( 
\mathrm{\lambda}_{n}^{3}\mathrm{\lambda}_{m}^{4}-\mathrm{\lambda}_{m}^{3}%
\mathrm{\lambda}_{n}^{4}\right) $ &  \\ 
$\mathcal{B}_{nm}$ & $=$ & $\left( \mathrm{\lambda}_{m}^{1}\mathrm{\lambda }%
_{n}^{1}+\mathrm{\lambda}_{m}^{2}\mathrm{\lambda}_{n}^{2}\right) -\left( 
\mathrm{\lambda}_{m}^{3}\mathrm{\lambda}_{n}^{3}+\mathrm{\lambda}_{m}^{4}%
\mathrm{\lambda}_{n}^{4}\right) $ & 
\end{tabular}%
\end{equation}
A possible solution is given by those wave vectors $\mathbf{K}$ obeying the
relation $\mathbf{K}.\mathrm{\beta}_{mn}=\frac{2}{d\sqrt{5}}\arctan\left( 
\mathcal{B}_{nm}/\mathcal{A}_{nm}\right) $.

\section{Re-deriving BC fermions}

In this section, we give the link between the above study based on SU$\left(
5\right) $ symmetry and the so called Bori\c{c}i-Creutz (\emph{BC}) model
having two zero modes associated with the light quarks up and down of QCD.
Recall that one of the important things in lattice \emph{QCD} is the need to
have a fermion action with a Dirac operator $\mathcal{D}$ having two zero
modes at points $K$ and $K^{\prime}$ of the reciprocal space; so that they
could be interpreted as the two light quarks. From this view, one may ask%
\textrm{\footnote{%
we thank the referee for pointing out this question which allowed us to
exhibit the relationship between our approach and BC fermions; see \cite{DS}
for explicit details.}} whether there exists a link between the present
analysis and the \emph{BC} fermions \textrm{\cite{BC,CB}}. In answering this
question, we have found that the \emph{BC} model can be indeed recovered
from the analysis developed in this paper. In what follows, we give the main
lines of the derivation.

\subsection{More on lattice action (\protect\ref{AC})}

One of the interesting lessons we have learnt from the analysis developed in
the previous sections is that the lattice action for 4D hyperdiamond
fermions may generally be written like, 
\begin{equation}
\begin{tabular}{ll}
$\mathcal{S}$ $\sim$ & $\frac{i}{4a}\dsum \limits_{\mathbf{r}}\left( \dsum
\limits_{l=0}^{4}\bar{\Psi}_{\mathbf{r}}\Gamma^{l}\Psi_{\mathbf{r+}a\mathbf{%
\lambda}_{l}}+\dsum \limits_{l=0}^{4}\bar{\Psi}_{\mathbf{r}}\bar{\Gamma}%
^{l}\Psi_{\mathbf{r-}a\mathbf{\lambda}_{l}}\right) ,$%
\end{tabular}
\label{SS}
\end{equation}
where $a=d\frac{\sqrt{5}}{2}$, the weight vectors $\mathbf{\lambda}_{l}$ as
in eqs(\ref{Z2}, \ref{Z9}) and where $\Gamma^{l}$ and their complex adjoints 
$\bar{\Gamma}^{l}$ are $4\times4$ complex matrices given by linear
combinations of the Dirac matrices $\mathrm{\gamma}^{\mu}$ as follows%
\begin{equation}
\begin{tabular}{llll}
$\Gamma^{l}=\left( \dsum \limits_{\mu=1}^{4}\mathrm{\gamma}^{\mu}\mathrm{%
\Omega}_{\mu}^{l}\right) $ & , & $\bar{\Gamma }^{l}=\left( \dsum
\limits_{\mu=1}^{4}\mathrm{\gamma}^{\mu}\mathrm{\bar{\Omega}}%
_{\mu}^{l}\right) $ & ,%
\end{tabular}
\label{GG}
\end{equation}
with $\Omega_{\mu}^{l}$ linking the lattice euclidian space time index $\mu$
and the index $l$ of the 5-dimensional representation of the $SU\left(
5\right) $ symmetry of the hyperdiamond. As such the lattice action (\ref{SS}%
) depends on the coefficients $\Omega_{\mu}^{l}$ capturing \emph{20} complex
numbers that form a $5\times4$ matrix representing the bi-fundamental of $%
SO\left( 4\right) \times SU\left( 5\right) $ 
\begin{equation}
\Omega_{\mu}^{l}=\left( 
\begin{array}{ccccc}
\Omega_{1}^{0} & \Omega_{1}^{1} & \Omega_{1}^{2} & \Omega_{1}^{3} & \Omega
_{1}^{4} \\ 
\Omega_{2}^{0} & \Omega_{2}^{1} & \Omega_{2}^{2} & \Omega_{2}^{3} & \Omega
_{2}^{4} \\ 
\Omega_{3}^{0} & \Omega_{3}^{1} & \Omega_{3}^{2} & \Omega_{3}^{3} & \Omega
_{3}^{4} \\ 
\Omega_{4}^{0} & \Omega_{4}^{1} & \Omega_{4}^{2} & \Omega_{4}^{3} & \Omega
_{4}^{4}%
\end{array}
\right) .
\end{equation}
This rank two tensor, which we decompose as $\left( \omega_{\mu},\Omega_{\mu
}^{\nu}\right) $ with $\omega_{\mu}=\Omega_{\mu}^{0}$ a complex 4 component
vector and $\Omega_{\mu}^{\nu}$ a complex $4\times4$ matrix, gives enough
freedom to engineer Dirac operators with a definite number of zero modes.
Below, we derive the constraint equations for the zero modes of the Dirac
operator; and in next subsection we apply the analysis to the \emph{BC}
model.

\subsubsection{Dirac operator}

In the reciprocal space, the lattice action (\ref{SS}) reads as%
\begin{equation}
\begin{tabular}{ll}
$\mathcal{S}$ $\sim$ & $\dsum \limits_{\mathbf{k}}\left( \dsum
\limits_{\mu=1}^{4}\bar{\Psi}_{\mathbf{k}}\mathcal{D}\Psi_{\mathbf{k}%
}\right) $%
\end{tabular}%
\end{equation}
with Dirac operator reading as follows 
\begin{equation}
\mathcal{D}=\frac{i}{4a}\dsum \limits_{\mu=1}^{4}\mathrm{\gamma}^{\mu}\left(
D_{\mu}+\bar{D}_{\mu}\right) ,   \label{DI}
\end{equation}
and where $D_{\mu}$ and its complex adjoint $\bar{D}_{\mu}$ are given by: 
\begin{equation}
\begin{tabular}{llll}
$D_{\mu}=\dsum \limits_{l=0}^{4}\mathrm{\Omega}_{\mu}^{l}e^{ia\mathbf{k}.%
\mathbf{\lambda}_{l}}$ & , & $\bar {D}_{\mu}=\dsum \limits_{l=0}^{4}\mathrm{%
\bar{\Omega}}_{\mu}^{l}e^{-ia\mathbf{k}.\mathbf{\lambda}_{l}}$ & .%
\end{tabular}
\label{ID}
\end{equation}
These operators depend on $40=2\left( 4+16\right) $ real numbers%
\begin{equation}
\begin{tabular}{llll}
$\omega_{\mu}=\frac{1}{2}\left( u_{\mu}+iv_{\mu}\right) $ & , & $\Omega
_{\mu}^{\nu}=\frac{1}{2}R_{\mu}^{\nu}+\frac{i}{2}J_{\mu}^{\nu}$ & ,%
\end{tabular}
\label{UV}
\end{equation}
and also on the five momenta $p_{l}=\hbar k_{l}$ along the $\mathbf{\lambda }%
_{l}$- directions. Since $k_{l}=\mathbf{k}.\mathbf{\lambda}_{l}$ and because
of SU$\left( 5\right) $ symmetry we have moreover the constraint relation 
\begin{equation}
\begin{tabular}{ll}
$k_{0}+k_{1}+k_{2}+k_{3}+k_{4}=0,\qquad\func{mod}\frac{2\pi}{a}$ & ,%
\end{tabular}
\label{F}
\end{equation}
allowing to express one of the five $k_{l}$'s in terms of the four others.\
For instance, we can express $k_{0}$ as follows: 
\begin{equation}
k_{0}=-\left( k_{1}+k_{2}+k_{3}+k_{4}\right) ,\qquad\func{mod}\frac{2\pi}{a}%
.   \label{G}
\end{equation}
The next step is to find the set of the wave vectors $k_{\mu}=\left(
k_{1},k_{2},k_{3},k_{4}\right) $ that give the zeros of the Dirac operator.
These zeros depend on the numbers $u_{\mu},$ $v_{\mu},$ $R_{\mu}^{\nu}$ and $%
J_{\mu}^{\nu}$ which can be tuned in order to get the desired number of
zeros.

\subsubsection{Zeros modes}

The zero modes of the Dirac operator $\mathcal{D}$ given by eqs(\ref{DI}-\ref%
{ID}) are obtained by solving the following constraint equations%
\begin{equation}
\dsum \limits_{\mu=1}^{4}\dsum \limits_{l=0}^{4}\mathrm{\gamma}^{\mu}\left( 
\mathrm{\Omega}_{\mu}^{l}+\mathrm{\bar{\Omega}}_{\mu}^{l}\right) \cos
ak_{l}+i\dsum \limits_{\mu=1}^{4}\dsum \limits_{l=0}^{4}\mathrm{\gamma}%
^{\mu}\left( \mathrm{\Omega}_{\mu}^{l}-\mathrm{\bar{\Omega}}%
_{\mu}^{l}\right) \sin ak_{l}=0,   \label{CS}
\end{equation}
together with the constraint eq(\ref{F}). Using the decomposition $\Omega
_{\mu}^{l}=\left( \omega_{\mu},\Omega_{\mu}^{\nu}\right) $, we can decompose
these constraints as follows%
\begin{equation}
\Lambda+\dsum \limits_{\mu=1}^{4}\left( \dsum \limits_{\nu=1}^{4}\mathrm{%
\gamma}^{\mu}\left( \mathrm{\Omega}_{\mu}^{\nu}+\mathrm{\bar{\Omega}}%
_{\mu}^{\nu}\right) \cos ak_{\nu}+i\dsum \limits_{\nu=1}^{4}\mathrm{\gamma}%
^{\mu}\left( \mathrm{\Omega}_{\mu}^{\nu}-\mathrm{\bar{\Omega}}%
_{\mu}^{\nu}\right) \sin ak_{\nu}\right) =0,
\end{equation}
where we have set%
\begin{equation}
\Lambda=\cos ak_{0}\dsum \limits_{\mu=1}^{4}\mathrm{\gamma}^{\mu}\left( 
\mathrm{\omega}_{\mu}+\mathrm{\bar{\omega}}_{\mu }\right) +i\sin ak_{0}\dsum
\limits_{\mu=1}^{4}\mathrm{\gamma}^{\mu}\left( \mathrm{\omega}_{\mu}-\mathrm{%
\bar{\omega}}_{\mu }\right) .
\end{equation}
Moreover using (\ref{UV}) we can put the above constraint relations into the
following equivalent form%
\begin{equation}
\Lambda+\dsum \limits_{\mu=1}^{4}\left( \dsum \limits_{\nu=1}^{4}\mathrm{%
\gamma}^{\mu}R_{\mu}^{\nu}\cos ak_{\nu}-\dsum \limits_{\nu=1}^{4}\mathrm{%
\gamma}^{\mu}J_{\mu}^{\nu}\sin ak_{\nu}\right) =0,   \label{13}
\end{equation}
and%
\begin{equation}
\Lambda=\cos ak_{0}\left( \dsum \limits_{\mu=1}^{4}\mathrm{\gamma}%
^{\mu}u_{\mu}\right) -\sin ak_{0}\left( \dsum \limits_{\mu=1}^{4}\mathrm{%
\gamma}^{\mu}v_{\mu}\right) =0,   \label{14}
\end{equation}
with $k_{0}$ given by eq(\ref{F}). Eqs(\ref{13}-\ref{14}) define a highly
non linear system of coupled equations in the four $k_{\nu}$'s; and are
difficult to solve in the generic case. To overcome this difficulty, one may
deal with these equations by focusing on adequate solutions for the $k_{v}$%
's; and engineer the corresponding $\Omega_{\mu}^{l}$ tensor. Below, we
apply this idea to the \emph{BC} model.

\subsection{\emph{BC} fermions}

\subsubsection{Deriving the model}

Bori\c{c}i-Creutz model \cite{BC} is a simple lattice QCD fermions for
modeling and simulating the interacting dynamics of the two light quarks up
and down. The Dirac operator of this model reads in the reciprocal space as
follows,%
\begin{equation}
\begin{tabular}{ll}
$\mathcal{D}_{BC}\text{ }\sim\text{ }\frac{i}{a}\dsum \limits_{\mu=1}^{4}%
\mathrm{\gamma}^{\mu}\sin ak_{\mu}-\frac{i}{a}\dsum \limits_{\mu=1}^{4}%
\mathrm{\gamma}^{\mu}\cos ak_{\mu}+\frac{i}{a}\dsum
\limits_{\mu=1}^{4}\Gamma\cos ak_{\mu}-\frac{2i}{a}\Gamma$ & ,%
\end{tabular}%
\end{equation}
with $\Gamma=\frac{1}{2}\left( \gamma^{1}+\gamma^{2}+\gamma^{3}+\gamma
^{4}\right) $. From this expression, one can check that this operator has
two zero modes given by the two following wave vectors,%
\begin{equation}
\begin{tabular}{lllll}
$\left( 1\right) $ & : & $\left( k_{1},k_{2},k_{3},k_{4}\right) $ & $=\left(
0,0,0,0\right) $ & , \\ 
$\left( 2\right) $ & : & $\left( k_{1},k_{2},k_{3},k_{4}\right) $ & $=\left( 
\frac{\pi}{2a},\frac{\pi}{2a},\frac{\pi}{2a},\frac{\pi}{2a}\right) $ & ,%
\end{tabular}%
\end{equation}
satisfying the remarkable property 
\begin{equation}
k_{1}+k_{2}+k_{3}+k_{4}=\frac{2\pi}{a},\quad\func{mod}\frac{2\pi}{a}. 
\label{P}
\end{equation}
Clearly the operator $\mathcal{D}_{BC}$ corresponds to a particular
configuration of the complex tensor $\Omega_{\mu}^{\nu}$ and the vector $%
\omega_{\mu}$. To see that is indeed the case, notice first that the matrix $%
\Gamma$ can be conveniently rewritten as $\Gamma=\frac{1}{2}\vartheta_{\mu
}\gamma^{\mu}$ with,%
\begin{equation}
\begin{tabular}{llll}
$\vartheta_{\mu}$ & $=$ & $\left( 1,1,1,1\right) $ & .%
\end{tabular}
\label{T}
\end{equation}
The same feature is valid for the sum $\sum_{\nu=1}^{4}\cos ak_{\nu}$ which
can be also put in the form $\sum_{\nu=1}^{4}\vartheta^{\nu}\cos ak_{\nu}$.
Putting these expressions back into the above $\mathcal{D}_{BC}$ relation,
we get: 
\begin{equation}
\mathcal{D}_{BC}\text{ }\sim\text{ }\frac{i}{a}\dsum \limits_{\mu,\nu=1}^{4}%
\mathrm{\gamma}^{\mu}\delta_{\mu}^{\nu}\sin ak_{\nu}-\frac{i}{a}\dsum
\limits_{\mu,\nu=1}^{4}\mathrm{\gamma}^{\mu}M_{\mu}^{\nu}\cos ak_{\nu}-\frac{%
2i}{a}\Gamma,   \label{DD}
\end{equation}
with 
\begin{equation}
M_{\mu}^{\nu}=\delta_{\mu}^{\nu}-\frac{1}{2}\vartheta_{\mu}\vartheta^{\nu}
\end{equation}
or more explicitly,%
\begin{equation}
M_{\mu}^{\nu}=\left( 
\begin{array}{cccc}
+\frac{1}{2} & -\frac{1}{2} & -\frac{1}{2} & -\frac{1}{2} \\ 
-\frac{1}{2} & +\frac{1}{2} & -\frac{1}{2} & -\frac{1}{2} \\ 
-\frac{1}{2} & -\frac{1}{2} & +\frac{1}{2} & -\frac{1}{2} \\ 
-\frac{1}{2} & -\frac{1}{2} & -\frac{1}{2} & +\frac{1}{2}%
\end{array}
\right) .
\end{equation}
Now comparing eq(\ref{DD}) with the general form of the Dirac operator of eq(%
\ref{DI}-\ref{ID}) which also reads like%
\begin{equation}
\Lambda+\dsum \limits_{\mu=1}^{4}\left( \dsum \limits_{\nu=1}^{4}\mathrm{%
\gamma}^{\mu}R_{\mu}^{\nu}\cos ak_{\nu}-\dsum \limits_{\nu=1}^{4}\mathrm{%
\gamma}^{\mu}J_{\mu}^{\nu}\sin ak_{\nu}\right) =0,
\end{equation}
we see that $\mathcal{D}_{BC}$ can be recovered by taking,%
\begin{equation}
\begin{tabular}{llll}
$R_{\mu}^{\nu}=-M_{\mu}^{\nu}$ & , & $J_{\mu}^{\nu}=-\delta_{\mu}^{\nu}$ & ,%
\end{tabular}
\label{R}
\end{equation}
and%
\begin{equation}
\Lambda=-2\Gamma=-\left( \gamma^{1}+\gamma^{2}+\gamma^{3}+\gamma^{4}\right)
.   \label{S}
\end{equation}
Eq(\ref{R}) leads to $\Omega_{\mu}^{\nu}=-\frac{1}{2}\left( M_{\mu}^{\nu
}+i\delta_{\mu}^{\nu}\right) $; by substituting $M_{\mu}^{\nu}$ by its
expression given above, the tensor $\Omega_{\mu}^{\nu}$ reads more
explicitly like 
\begin{equation}
\Omega_{\mu}^{\nu}=-\frac{\left( 1+i\right) }{2}\delta_{\mu}^{\nu}+\frac {1}{%
4}\vartheta_{\mu}\vartheta^{\nu}.
\end{equation}
The second constraint relation (\ref{S}) requires 
\begin{equation}
\left( u_{\mu}\cos ak_{0}-v_{\mu}\sin ak_{0}\right) =-\vartheta_{\mu},
\end{equation}
with $\vartheta_{\mu}$\ as in (\ref{T}). Moreover, using eqs(\ref{P},\ref{G}%
), we end with%
\begin{equation}
u_{\mu}=-\vartheta_{\mu},
\end{equation}
and $v_{\mu}$ a free vector which, for simplicity, we set to zero. Thus the
tensor $\Omega_{\mu}^{l}=\left( \omega_{\mu},\Omega_{\mu}^{\nu}\right) $
describing the BC fermions is given by%
\begin{equation}
\Omega_{\mu}^{l}=\left( 
\begin{array}{ccccc}
-\frac{1}{2} & -\frac{1+2i}{4} & \frac{1}{4} & \frac{1}{4} & \frac{1}{4} \\ 
-\frac{1}{2} & \frac{1}{4} & -\frac{1+2i}{4} & \frac{1}{4} & \frac{1}{4} \\ 
-\frac{1}{2} & \frac{1}{4} & \frac{1}{4} & -\frac{1+2i}{4} & \frac{1}{4} \\ 
-\frac{1}{2} & \frac{1}{4} & \frac{1}{4} & \frac{1}{4} & -\frac{1+2i}{4}%
\end{array}
\right) ,   \label{26}
\end{equation}
with the trace property $\sum_{l}\Omega_{\mu}^{l}=-\frac{i}{2}\vartheta_{\mu}
$.

\subsubsection{Symmetries}

Here we want to make a comment on particular symmetries of \emph{4D} lattice 
\emph{QCD} fermions by following the analysis of ref.\textrm{\cite{BB2}
where the study of the }renormalization of this class of models has been
explicitly done. There, it has been found that the breaking of discrete
symmetries, such as parity $\mathcal{P}$ : $\Psi(\vec{k},k_{4})\rightarrow$ $%
\mathrm{\gamma }_{4}\Psi(-\vec{k},k_{4})$ and time- reversal $\mathcal{T}$: $%
(\vec{k},k_{4})\rightarrow$ $\mathrm{\gamma}_{5}\mathrm{\gamma}_{4}\Psi(\vec{%
k},-k_{4})$, is behind the appearance of relevant dimension 3 operators $%
\mathcal{O}_{3}^{\left( i\right) }$ and marginal dimension 4 ones $\mathcal{O%
}_{4}^{\left( j\right) }$ in the analysis of the Symanzik effective theory
with lagrangian $\mathcal{L}_{eff}=\frac{1}{a^{4}}\sum
_{n}a^{n}\sum_{j}c_{n}^{\left( j\right) }\mathcal{O}_{n}^{\left( j\right) }$%
. Following the above mentioned work, one starts from the \emph{4D} lattice
action, 
\begin{equation}
\mathcal{S}\sim\frac{1}{2}\sum_{\mathbf{x}}\dsum \limits_{\mu=1}^{4}\left[
\Psi_{\mathbf{x}}^{+}\text{ }\boldsymbol{A}^{\mu}\text{ }\Psi_{\mathbf{x}+a%
\mathbf{\upsilon}_{\mu}}-\Psi_{\mathbf{x}}^{+}\text{ }\boldsymbol{\bar{A}}%
^{\mu}\text{ }\Psi_{\mathbf{x}-a\mathbf{\upsilon}_{\mu}}-2iBC\text{ }\Psi_{%
\mathbf{x}}^{+}\text{ }\mathrm{\gamma}^{4}\text{ }\Psi_{\mathbf{x}}\right] . 
\label{B2}
\end{equation}
which depends on two real parameters $B$ and $C$ that are fixed by physical
requirements and symmetries. This typical action depends also on particular
combinations of gamma matrices $\boldsymbol{A}^{\mu}=\sum_{\nu=1}^{4}\mathrm{%
\gamma}^{\nu}A_{\nu}^{\mu}$ where the coefficients $A_{\nu}^{\mu}$, given in 
\textrm{\cite{BB2}, }form an invertible $4\times4$ matrix with\textrm{\ }$%
\det\left( A_{\nu}^{\mu}\right) =-16iB$. Notice that setting $B=1$, $C=\frac{%
\sqrt{2}}{2}$ one recovers the Bori\c{c}i action. By performing
transformations of (\ref{B2}) using Fourier integrals to move to the
reciprocal space, similarity operations to exhibit particular symmetries,
expansion in powers of the lattice spacing parameter $a$ to use the Symanzik
effective theory; and switching on the usual gauge interactions $\partial
_{\mu}\rightarrow\mathfrak{D}_{\mu}=\partial_{\mu}-ig\mathcal{A}_{\mu}$ with
field strength $\mathcal{F}_{\mu\nu}=\frac{i}{g}\left[ \mathfrak{D}_{\mu },%
\mathfrak{D}_{\nu}\right] $, we get up to the first order in the parameter $a
$ the following effective field action,%
\begin{equation*}
\mathcal{L}_{eff}=\dsum \limits_{\mathbf{x}}\left[ \bar{Q}\left(
\gamma^{\mu}\otimes I\right) \mathfrak{D}_{\mu}Q-\frac{1}{4}\mathcal{F}%
_{\mu\nu}\mathcal{F}^{\mu\nu}+a\mathcal{O}_{5}+ord\left( a^{2}\right) \right]
, 
\end{equation*}
with $Q_{\alpha}$ standing for the quark isodoublet $\left( u_{\alpha
},d_{\alpha}\right) $ and $\mathcal{O}_{5}$ some dimension 5 operator that
can be found in \textrm{\cite{BB2}}. This effective theory has several
symmetries in particular: (\textbf{1}) the manifest gauge invariance, (%
\textbf{2}) $U_{B}\left( 1\right) $ baryon number, (\textbf{3}) $U_{L}\left(
1\right) \times U_{R}\left( 1\right) $ chiral symmetry, (\textbf{4}) $%
\mathcal{CPT}$ invariance and (\textbf{5}) symmetry under $S_{4}$
permutation of the four hyperplane axis corresponding to $C=BS$ with $%
e^{iK}=C+iS$. The authors of \textrm{\cite{BB2} }concluded their work by two
remarkable results: (\textbf{i}) the engineering of chirally symmetric
action with minimal fermion doubling which does not generate dimension three
operators $\mathcal{O}_{3}$ is possible as far as $\mathcal{PT}$ symmetry is
preserved. This invariance is sufficient to forbid the relevant dimension 3
operators $\mathcal{O}_{3}^{\left( i\right) }$ whose typical forms are
listed below, 
\begin{equation*}
\begin{tabular}{lllll}
broken $\mathcal{P}$ & $:$ & $\mathcal{O}_{3}^{\left( 1\right) }=i$ $\bar{%
\Psi}_{\vec{k},k_{4}}\text{ }\gamma_{j}$ $\Psi_{\vec{k},k_{4}}$ & , & $%
\mathcal{O}_{3}^{\left( 2\right) }=i$ $\bar{\Psi}_{\vec{k},k_{4}}\text{ }%
\gamma_{4}\gamma_{5}$ $\Psi_{\vec{k},k_{4}}$ \\ 
&  &  &  &  \\ 
broken $\mathcal{T}$ & $:$ & $\mathcal{O}_{3}^{\left( 3\right) }=i$ $\bar{%
\Psi}_{\vec{k},k_{4}}\text{ }\gamma_{j}\gamma_{5}$ $\Psi_{\vec{k},k_{4}}$ & ,
& $\mathcal{O}_{3}^{\left( 4\right) }=i$ $\bar{\Psi}_{\vec{k},k_{4}}$ $%
\gamma_{j}\gamma_{5}$ $\Psi_{\vec{k},k_{4}}$ \\ 
&  &  &  & 
\end{tabular}
\end{equation*}
with $\gamma_{j}$ standing for $\gamma_{1}$, $\gamma_{2}$ , $\gamma_{3}$. (%
\textbf{ii}) For particular values of parameters of the theory, there may
emerge some additional non standard symmetries which could be used to
eliminate the relevant operators. These results are important and may serve
as guide lines in dealing with this problem by using the hyperdiamond
symmetries based on roots and weights of $SU\left( 5\right) $. Below, we
give a comment on this matter; an exact answer needs however a deeper
analysis. In the $SU\left( 5\right) $ framework, the previous action (\ref%
{B2}) gets extended as follows%
\begin{equation}
\mathcal{S}_{su_{5}}\sim\frac{i}{a}\dsum \limits_{\mathbf{x}}\dsum
\limits_{\mu}\left( \dsum \limits_{l=0}^{4}\left[ \bar{\Psi}_{\mathbf{x}}%
\mathrm{\gamma}^{\mu}\mathrm{\Omega}_{\mu}^{l}\Psi_{\mathbf{x+}a\mathbf{%
\lambda}_{l}}+\bar{\Psi}_{\mathbf{x}}\mathrm{\gamma}^{\mu}\mathrm{\bar{\Omega%
}}_{\mu}^{l}\Psi_{\mathbf{x-}a\mathbf{\lambda}_{l}}\right] \right) ,
\end{equation}
where $\mathrm{\Omega}_{\mu}^{l}$ as before and the $\mathbf{\lambda}_{l}$'s
are the weight vectors of the 5-dimensional representation of SU$\left(
5\right) $. Clearly, this lattice action is more general than eq(\ref{B2});
and has two interesting features that are useful in dealing with the study
of underlying symmetries and renormalization of $\mathcal{S}_{su_{5}}$.
First, the $SU\left( 5\right) $ property (\ref{LA}) on the weight vectors
namely $\sum_{l}\mathbf{\lambda}_{l}^{\mu}=0$ induces in turns the following
constraint relation on the wave vectors $k_{\mu},$%
\begin{equation*}
\begin{tabular}{lll}
$\dsum \limits_{l=0}^{5}k_{l}=0$, & with & $k_{l}=\dsum
\limits_{\mu=1}^{5}k_{\mu}.\lambda_{l}^{\mu}$.%
\end{tabular}
\end{equation*}
This constraint is invariant under $\mathcal{PT}$ symmetry acting on wave
vectors as $k_{\mu}\rightarrow-k_{\mu}$; but not preserved under parity $%
\mathcal{P}$ nor time- reversal $\mathcal{T}$ separately. Second the
generalized action $\mathcal{S}_{su_{5}}$ depends on \emph{20} complex (%
\emph{40} real) moduli carried by the tensor $\mathrm{\Omega}_{\mu}^{l}$.
This number gives quite enough freedom to engineer QCD-like models with two
zeros for the Dirac operator as we have done in case of \emph{BC} model; may
lead to desired symmetries of the Symanzik effective theory that follow from
the expansion of the action $\mathcal{S}_{su_{5}}$ in powers of the lattice
parameter; and may allow to make appropriate choices to eliminate relevant
operators. Progress in this matter will be reported in a future occasion.\ \ 

\section{Conclusion}

In this paper, we have studied the lattice fermion action for pristine \emph{%
4D} hyperdiamond $\mathcal{H}_{4}$ with desired properties for \emph{4D}
lattice QCD simulations. Using the $SU\left( 5\right) $ hidden symmetry of $%
\mathcal{H}_{4}$, we have constructed a BBTW- like lattice model by
mimicking \emph{2D} graphene model. To that purpose, we first studied the
link between the construction of \cite{B2} and $SU\left( 5\right) $; then we
refined BBTW lattice action by using the weight vectors $\mathbf{\lambda}%
_{0},$ $\mathbf{\lambda}_{1},$ $\mathbf{\lambda}_{2},$ $\mathbf{\lambda}_{3},
$ $\mathbf{\lambda}_{4}$ of the \emph{5}-dimensional representation of $%
SU\left( 5\right) $. After that we studied explicitly the solutions of the
zeros of the Dirac operator in terms of the $SU\left( 5\right) $ simple
roots $\mathbf{\alpha}_{1},$ $\mathbf{\alpha}_{2},$ $\mathbf{\alpha}_{3},$ $%
\mathbf{\alpha}_{4}$, and its fundamental weights $\mathbf{\omega}_{1},$ $%
\mathbf{\omega}_{2},$ $\mathbf{\omega}_{3},$ $\mathbf{\omega}_{4}$,. We have
found that the zeros of the Dirac operator live at the sites $\mathbf{k=}%
\frac{4\pi}{d\sqrt{5}}\left( N_{1}\mathbf{\omega}_{1}+N_{2}\mathbf{\omega }%
_{2}+N_{3}\mathbf{\omega}_{3}+N_{4}\mathbf{\omega}_{4}\right) $ of the
reciprocal lattice $\mathcal{H}_{4}^{\ast}$; with $N_{i}$ integers. In
addition to their quite similar continuum limit, we have also studied the
link between the Dirac operator following from our construction and the one
suggested by Creutz using quaternions; the Dirac operator in our approach
may be viewed as a \emph{"complexification"} of the Creutz one where the
role played by the $\sin\left( p_{i}\right) $'s and the $\cos\left(
p_{i}\right) $'s is now played by $e^{ip_{i}}$ as shown in eqs(\ref{te}) and
(\ref{et}). The exact link between our approach and the Bori\c{c}i-Creutz
fermions has been worked out with details in section 6; where it is shown
that the \emph{BC} action follows exactly from (\ref{SS}) with eqs(\ref{GG},%
\ref{26}) giving the linear combinations of the Dirac matrices of the model.%
\newline
It is also interesting to notice that our approach is general; and applies
straightforwardly to lattice systems in diverse dimensions. The fact that
the \emph{4D} hyperdiamond is related to $SU\left( 5\right) $ fundamental
weights $\mathbf{\omega}_{1},$ $\mathbf{\omega}_{2},$ $\mathbf{\omega}_{3},$ 
$\mathbf{\omega}_{4}$, and its simple roots $\mathbf{\alpha}_{1},$ $\mathbf{%
\alpha}_{2},$ $\mathbf{\alpha}_{3},$ $\mathbf{\alpha}_{4}$ is not specific
for 4-dimensions; it can be extended to generic dimensions \emph{D} where
the underlying D-dimensional hyperdiamond lattice has a hidden $SU\left(
D+1\right) $ symmetry with simple roots $\mathbf{\alpha}_{1},$...$, $ $%
\mathbf{\alpha}_{D}$ and fundamental weights $\mathbf{\omega}_{1},\mathbf{...%
},$ $\mathbf{\omega}_{D}$. From this view, the 2D graphene has therefore a
hidden $SU\left( 3\right) $ symmetry as reported in details in \cite{A3}.
Our construction applies as well to the fermion actions given in \cite{B5}.%
\newline

\textbf{Acknowledgement} :\newline
L.B Drissi would like to thank ICTP for associate membership and E.H Saidi
thanks URAC-09, CNRST.

\end{document}